# Visible-frequency hyperbolic plasmon polaritons in a natural van der Waals crystal

*Giacomo Venturi*[†], *Andrea Mancini*[†] *, *Nicola Melchioni*[†], *Stefano Chiodini, Antonio Ambrosio**

G. Venturi, A. Mancini, N. Melchioni, S. Chiodini, A. Ambrosio

Centre for Nano Science and Technology, Fondazione Istituto Italiano di Tecnologia, Via Rubattino 81, Milano 20134, Italy

G. Venturi

Physics Department, Politecnico di Milano, Piazza Leonardo Da Vinci 32, Milano 20133, Italy

[†] These authors contributed equally to this work

*Correspondence: andrea.mancini@iit.it; antonio.ambrosio@iit.it

**Abstract**

Controlling light at subwavelength scales is one of the main challenges of nanophotonics. Leveraging hyperbolic polaritons supporting arbitrarily large wavevectors can lead to extreme light confinement, effectively overcoming the diffraction limit. Hyperbolicity was initially realized in artificial metamaterials, but their performances are limited by high losses in the metallic components. While recently discovered natural low-loss hyperbolic phonon polaritons initiated a revival in the interest for hyperbolic materials, they are confined to the mid-infrared frequency range, limiting their use for several applications. Some hyperbolic materials at visible frequencies have been studied, but they are either very lossy or only feature out-of-plane hyperbolicity. Here, we demonstrate the presence of low-loss, in-plane hyperbolic plasmon polaritons in the visible and near-infrared in thin films of $MoOCl_2$, a natural van der Waals crystal. The polariton dispersion is predicted based on the framework of light propagation in biaxial media, and experimentally confirmed by real space nano imaging on exfoliated flakes. $MoOCl_2$ constitutes a novel material platform for visible-range applications leveraging the unboundedness of hyperbolic modes, such as hyperlensing, Purcell factor enhancement and super-resolution imaging, without the drawbacks of metamaterials.

## Introduction

In the past decade, polaritons[1] - mixed light-matter states formed by coupling photons with quantum excitations of crystals (e.g., optical phonons or plasmons) - have enabled the control of light at the subwavelength scale, emerging as game changers for several applications encompassing nanoscale heat generation[2,3], photocatalysis[4,5], enhanced surface spectroscopies[6,7], wavefront engineering[8–10] and nonlinear optics[11,12].

The confinement provided by polaritons is determined by the material dielectric permittivity and described by the isofrequency contours (IFCs) defining the allowed wavevectors for propagating modes at a certain wavelength. When the permittivity tensor is diagonal with all negative components, polaritons propagate at the material boundary with spherical or elliptic wavefronts, associated with closed IFCs (Fig. 1a). Conversely, anisotropic materials with a permittivity that is negative (i.e., metallic-like) and positive (i.e., dielectric-like) along orthogonal directions support polaritons with hyperbolic wavefronts, defined by strongly anisotropic IFCs featuring arbitrarily large wavevectors (Fig. 1b)[13,14]. Thus, as the wavelength is inversely proportional to the wavevector, hyperbolic polaritons can be leveraged to confine light at extremely subwavelength scales[15,16].

A versatile platform for polaritonics is provided by van der Waals (vdW) materials[1], such as hexagonal boron nitride (hBN)[17–21], α-$MoO_3$[22,23] and β-$Ga_2O_3$[24,25], recently shown to support hyperbolic phonon polaritons (PhPs). Their highly anisotropic lattice results in different phonon resonances along each crystal direction, causing the permittivity tensor elements to have opposite signs at mid-infrared (IR) frequencies (Fig. 1c). Exploiting the hyperbolic IFCs, highly controlled unidirectional[26] and canalized[27] PhPs propagation, as well as negative reflection[28] and refraction[29] have been recently demonstrated. While PhPs are promising for applications in thermal radiation engineering[30] and vibrational sensing[31], highly sought implementations for hyperlensing[32], enhanced quantum emission[33], super resolution imaging[34] and high-resolution photolithography[35] require operation in the visible range. As PhPs are limited to IR wavelengths ($\lambda_0 > 5$ μm) due to the characteristic energy of optical phonons, exploring alternative platforms is essential.

Operation at visible frequencies has been realized with nanostructured hybrid metal-dielectric materials, called metamaterials, which can be tailored to show hyperbolic plasmon polaritons (PPs)[14]. Even though negative refraction[36] and hyperlensing[32,37] have been demonstrated with hyperbolic metamaterials in the visible, such artificial materials are constrained by the limited fabrication resolution of lithography processes and can only work in the long wavelength limit for which the effective medium approximation holds[14], thus hindering their operation at large momenta. Moreover, the presence of noble metals in the metamaterial structure makes them typically very lossy at visible frequencies.

Natural hyperbolic materials can help circumvent such limitations. Hyperbolicity in the visible range has been found originating from anisotropic excitons[38,39], interband transitions[40], plasmons (Fig. 1d)[41,42], or a combination of them[43,44]. However, until now either the presence of high losses or the out-of-plane nature of the response prevented the observation of natural hyperbolic polaritons in the visible. Only recently, near-IR real space nanoimaging of hyperbolic exciton polaritons has been reported in CrSBr thin films[45], but only at cryogenic temperatures. Moreover, their coexistence with waveguides modes hampered their visualization. As a result, no clear evidence of room-temperature in-plane hyperbolic polaritons in the visible range has been reported so far.

In this work, we report the discovery of low-loss in-plane hyperbolic plasmon polaritons at visible and near-IR frequencies in thin films of the natural vdW crystal molybdenum oxide dichloride ($MoOCl_2$), whose in-plane crystal structure is sketched in Fig. 1e. First, we give a description of the PPs supported by bulk $MoOCl_2$ in the theoretical framework of polariton propagation in biaxial media. Our analysis reveals the simultaneous presence of hyperbolic PPs as well as directional “ghost” modes in the material (i.e., modes showing damped propagation even in the absence of material losses)[46], as represented by the IFCs in Fig. 1f. Thanks to the broad hyperbolic region of $MoOCl_2$, these modes span the near-IR and visible ranges (Fig. 1g). We then discuss the modification of the PPs IFCs from the bulk to the thin film limit. We experimentally confirm the optical anisotropy of $MoOCl_2$ through high-resolution atomic force microscopy and far-field reflectivity. Finally, we investigate the predicted thin-film IFCs through real space polariton nanoimaging on exfoliated $MoOCl_2$ (as illustrated in Fig. 1h). Our demonstration of low-loss in-plane natural hyperbolicity in the visible range paves the way for enhanced control of light confinement through the unboundedness of polariton IFCs, with promising applications in high-frequency 2D materials-based optoelectronics.

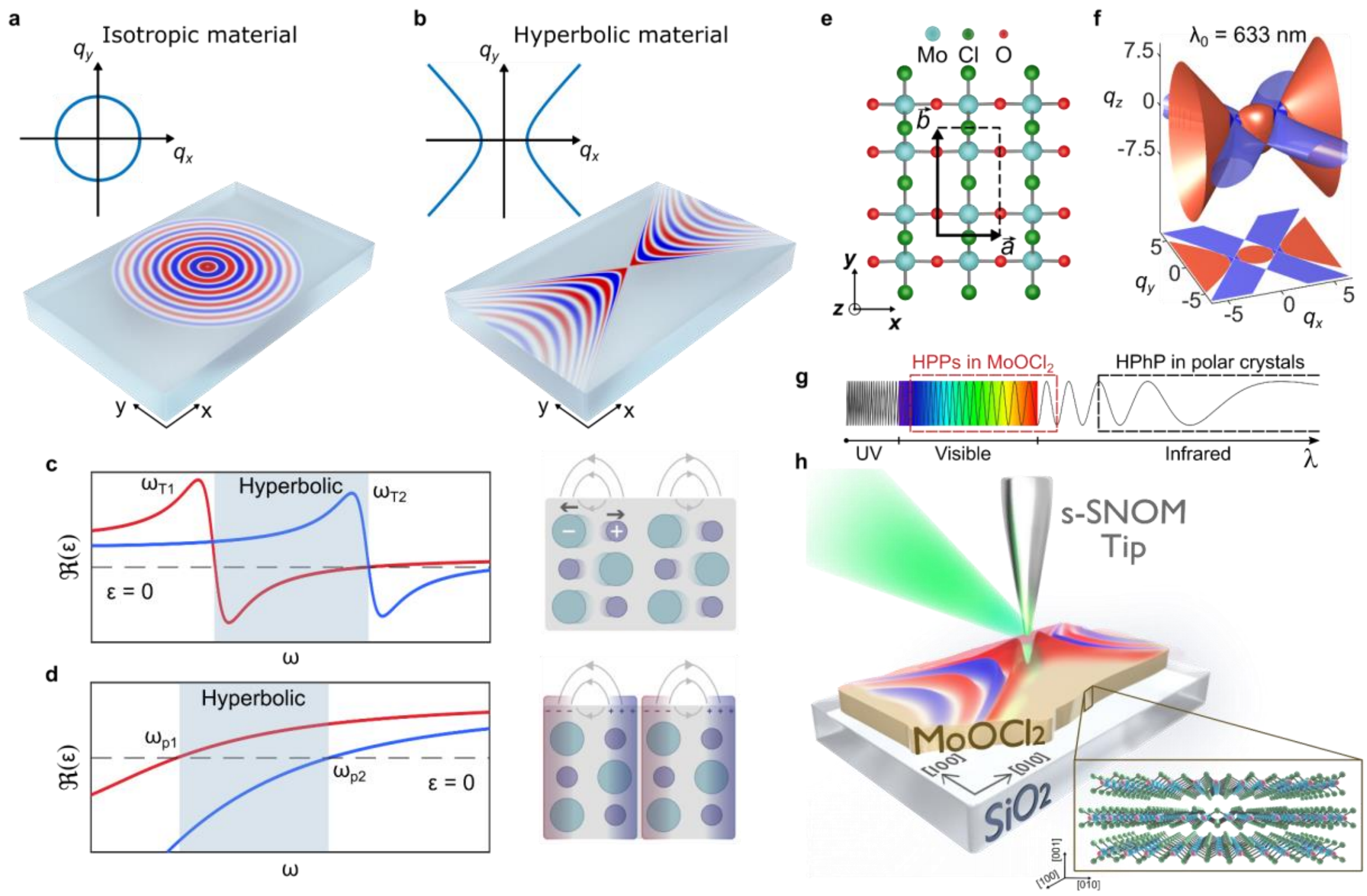

**Figure 1. Hyperbolic plasmon polaritons at visible frequencies. a** Polaritons in isotropic materials propagate with circular wavefronts, associated with closed IFCs (inset). **b** Polaritons in hyperbolic materials propagate with hyperbolic wavefronts set by the unbounded IFCs (inset). **c** Different (optical) phonon resonances ($\omega_{T1}$, $\omega_{T2}$) cause the dielectric permittivity to be negative in one direction, defining a hyperbolic spectral region (grey shaded area, left panel) in which PhPs are formed (right panel). **d** Different plasma frequencies ($\omega_{p1}$, $\omega_{p2}$) give rise to hyperbolic windows (grey shaded area, left panel), where light couples to quasi-free electrons forming PPs (right panel). **e** In-plane crystal structure of $MoOCl_2$, where 2D unit cell vectors are shown. **f** IFCs of bulk $MoOCl_2$ computed for $\lambda_0 = 633$ nm. **g** Hyperbolic PhPs lie in the IR region due to the limited energy of phonons. Hyperbolic PPs in $MoOCl_2$, instead, span a large portion of the visible and near IR. **h** $MoOCl_2$ PPs can be imaged with s-SNOM, consisting of a metallic tip illuminated by (visible) light. The scattered near-field provides the high-momenta required to excite sub-diffraction polaritons. Inset: crystal structure of bulk $MoOCl_2$.

## Main text

### Plasmon polaritons in bulk $MoOCl_2$

$MoOCl_2$ is a layered vdW material belonging to the subgroup of oxychlorides with formula $XOCl_2$ (X = Mo, Nb, Ta, V, Os), whose crystal structure is characterized by a monoclinic lattice (space group C2/$m$)[47,48]. The structure of a single layer of the $MoOCl_2$ crystal consists of a central plane made of Mo-O chains (extending in the [100] direction) sandwiched between two layers of Cl atoms as shown in the inset of Fig. 1h.[49] The in-plane lattice takes a rectangular shape (Fig. 1e), with orthogonal unit cell vectors $\vec{a}, \vec{b}$ with sizes 3.8 Å and 6.5 Å[49], respectively. Monolayers are stacked in the [001] direction at a distance of 12.7 Å to form the bulk.

Recent theoretical works derived the optical constants of biaxial $MoOCl_2$ [47,48]. The calculated dielectric tensor is diagonal with three different principal components $(\varepsilon_x, \varepsilon_y, \varepsilon_z)$ (here we associate the $(x, y, z)$ directions with the crystalline axes $[100], [010]$ and $[001]$). A broad hyperbolic region across the near-IR and visible ranges where $\varepsilon_x < 0$ and $\varepsilon_y, \varepsilon_z > 0$ is expected, originating from a metallic character with highly different values for the plasma frequency ($\omega_p$) along distinct crystalline directions (as sketched in Fig. 1d; for details refer to Supplementary Information S1). Intuitively, this is a result of the Mo-O atoms behaving as isolated chains due to their much stronger lattice binding compared to the other crystal directions[47]. The $MoOCl_2$ hyperbolic region is also predicted to be low-loss[48] (see

Supplementary Information S1), suggesting the possibility of observing the propagation of hyperbolic PPs.

Based on the theoretical values of the $MoOCl_2$ optical constants, we calculate its IFCs at $\lambda_0 = 633$ nm in the hyperbolic spectral range (Fig. 1f, where $\boldsymbol{q} = (q_x, q_y, q_z)$ is the normalized wavevector; see Supplementary Information S2).[50] The orange IFCs show the coexistence of hyperbolic modes with unbounded wavevectors in the $q_x$ direction, alongside a central spheroidal region resembling the IFC of a dielectric material. Furthermore, our calculations predict the existence of directional ghost modes (light purple)[46,51].

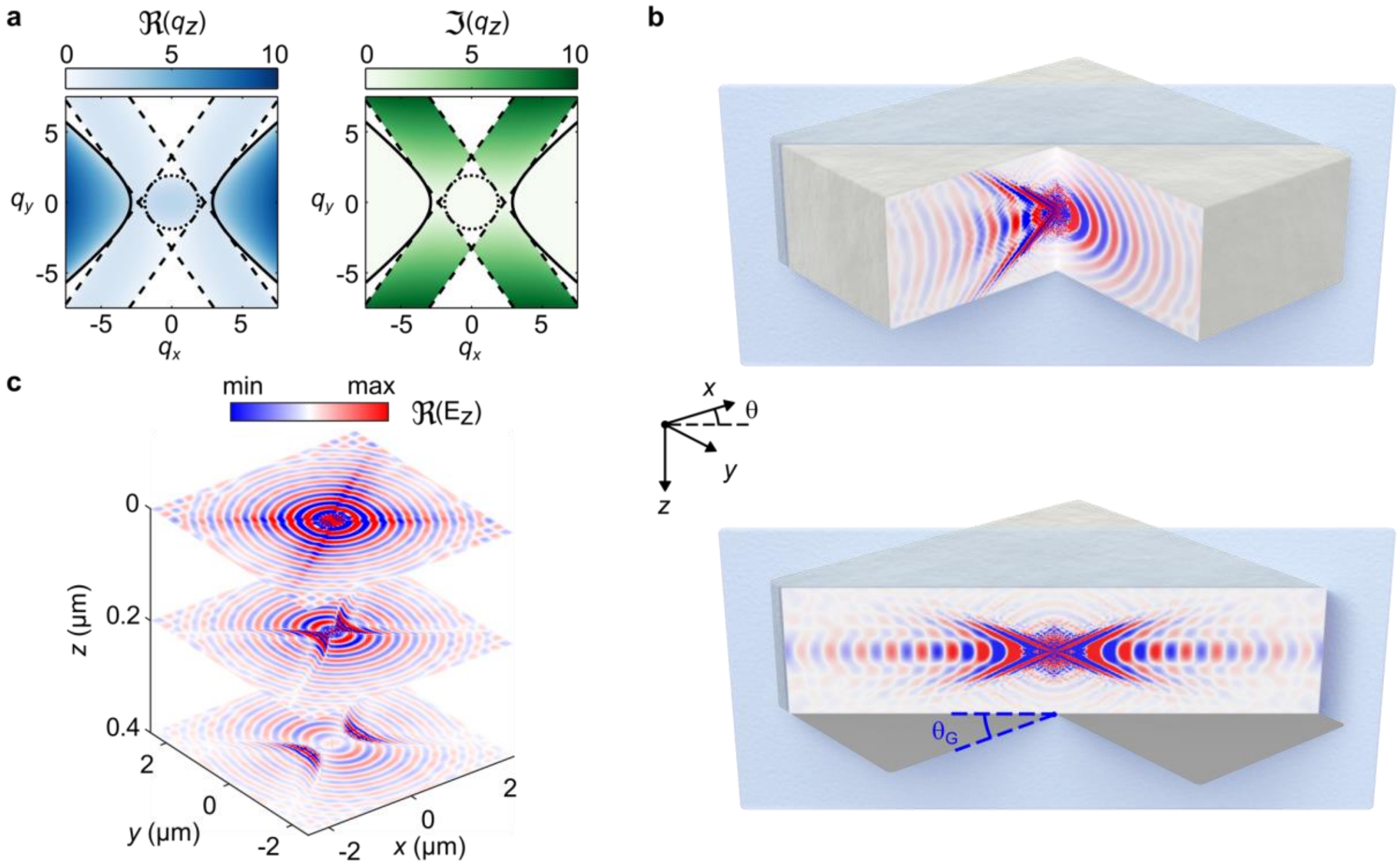

**Figure 2. Anisotropic plasmon polaritons propagation in bulk $MoOCl_2$. a** In-plane projection of the real (left) and imaginary (right) parts of the calculated IFCs ($\lambda_0 = 633$ nm, without material losses). Solid and dashed lines delimit the boundaries of the hyperbolic and ghost propagating regions. Dotted line corresponds to the $|\boldsymbol{q}| = n_z$ circle ($MoOCl_2$ light-cone). **b** Three-dimensional simulations of PPs emitted by a near-field source located in the middle of a slab of $MoOCl_2$ slab (1.5 $\mu m$ thickness). Projections of the $E_z$ field on the $x$-$z$ and $y$-$z$ planes (top) and on the $y$-$z$ plane rotated clockwise by $\theta_G = 36°$ around the $z$-axis, aligned with the directional propagation of ghost PPs. **c** In-plane values of the $E_z$ field in **b** at different distances from the near-field source.

Ghost modes are an exotic class of waves featuring damped propagation with a complex wavevector even in the absence of intrinsic material losses[51]. To investigate the presence of ghost modes it is therefore instructive to inspect the in-plane projections of the real (left) and imaginary (right) parts of the analytically derived IFCs shown in Fig. 2a ($\lambda_0 = 633$ nm). In the $(q_x, q_y)$ space, three distinct domains can be observed as delimited by the black continuous, dashed, and dotted lines, corresponding to three different propagating modes. Spherical-like waves exist within the dotted circle, whose radius is determined by the out-of-plane refractive index of $MoOCl_2$ ($|\boldsymbol{q}| = n_z$). The continuous lines delimit the hyperbolic regions asymptotically approaching straight lines defined by $q_y/q_x = \pm\sqrt{-\varepsilon_x/\varepsilon_y}$. The dashed curves delimit the region in momentum space where ghost modes exist, as this is the only part of the IFCs where at the same time both the real and imaginary part of $q_z$ are nonzero (even when the material is lossless)[51]. Similarly to hyperbolic PPs, ghost PPs are highly confined as their IFCs are unbounded. Ghost PPs are also strongly directional as defined by the angular coefficient of the straight IFCs boundaries $q_y/q_x = \pm\sqrt{(\varepsilon_x - \varepsilon_z)/(\varepsilon_y - \varepsilon_z)}$.[51] Remarkably, ghost modes in $MoOCl_2$ are fundamentally different from the ones recently observed in calcite [52]. In this latter case, ghost modes appear as surface waves at the boundary between an off-cut uniaxial crystal and a dielectric, and feature in-plane hyperbolic dispersion. In $MoOCl_2$, ghost waves are intrinsic bulk modes of a biaxial crystal, without the need of an interface with an off axis cut. Their existence only depends on the relative magnitude of the dielectric tensor components (see Supplementary Information S2 for more details) and they do not feature hyperbolic dispersion. They rather populate a cross-shaped region in momentum space (see Fig. 2a, b).

To verify the prediction of the analytical calculations, we perform three-dimensional electromagnetic simulations at $\lambda_0 = 633$ nm, the same free-space wavelength of the computed theoretical IFCs of Fig. 1e and Fig. 2a. The thickness of the slab is chosen so that there is no appreciable interaction between the polaritons and the top/bottom interfaces, effectively approximating the behaviour of an infinite crystal. By cutting the simulated $\mathrm{Re}(E_z)$ field distribution along orthogonal planes (top panel of Fig. 2b), we observe the expected propagation of hyperbolic PPs in the $x$ direction ($x$-$z$ cut), whereas along $y$ spherical wavefronts are in line with the dipolar emission in a dielectric environment ($y$-$z$ cut). The propagation direction for the observed PPs is determined by the Poynting vector, which is perpendicular to the IFCs[53]. For ghost waves, the angular coefficient of the straight lines defining the region of ghost modes is $54°$ at $\lambda_0 = 633$ nm (dashed lines in Fig. 2a), yielding a predicted in-plane propagation direction of $\theta_G = 36°$ in real-space. The fields evaluated in a plane rotated by $\approx \theta_G$ with respect to the $x$-$z$ plane (bottom panel of Fig. 2b) show the existence of highly directional PPs confined both in $z$ and in the $\theta_G$-oriented plane, which we identify with the ghost modes predicted by our theoretical analysis (see Supplementary Information S2).

In the $\theta_G$ cut, we also observe the presence of hyperbolic PPs with a smaller out-of-plane opening angle compared to the ones appearing in the $x$-$z$ cut (top panel of Fig. 2b). This effect is explained by considering the direction of the Poynting vector in relation to the orientation of the cut plane (see Supplementary Information S2).

The properties of PPs can be also visualized from the in-plane projections of $\mathrm{Re}(E_z)$ at three different depths (0, 200 and 400 nm from the near-field source plane, located at the centre of the slab) shown in Fig. 2c. Ghost PPs are confined at $z = 0$ and are canalized along four specific directions[54]. An almost isotropic emission (equal along $x$ and $y$, but different along $z$) associated with dielectric modes (central spheroid in the IFCs of Fig. 1f) is also present in the simulations, which interferes with both hyperbolic and ghost PPs.

**Plasmon polaritons in $MoOCl_2$ thin films**

Having described the bulk modes, we move to the analysis of PPs in thin films. As recently shown for hBN[17] and $MoO_3$[23], bulk hyperbolic polaritons get reflected at the material's top and bottom interfaces, as sketched in Fig. 3a, forming waveguide-like modes propagating inside the flake (see Supplementary Information S5). To better understand how the modes evolve with the film thickness, we report in Fig. 3b the in-plane dispersion calculated through the transfer matrix method for four different thicknesses between 80 nm and 20 nm. At 80 nm we observe two hyperbolic modes (symmetric with respect to the $q_y$ axis), lying outside the substrate isotropic light-cone, with two different apertures: the lowest order bulk mode, with higher wavevectors, and a mode with smaller curvature, only visible for thicknesses below $\sim 200$ nm (see Supplementary Information S5). The hyperbola of the latter does not actually close at its vertex and its intensity drops to zero for $q_x < n_z$. In this region, the IFCs change from a hyperbolic to a lenticular-like dispersion, in a similar fashion to reported cases of mid-IR surface PhPs modes[54]. Such a transition in the mode shape when crossing $n_z$ can be better appreciated for a $40$ nm-thick film, shown in the inset of Fig. 3b. By further reducing the thickness, the vertex of the hyperbola lies completely outside of the $n_z$ circle and the lenticular mode disappears as shown in the dispersion for the $20$ nm film. Lenticular modes in $MoOCl_2$ share some features with leaky modes observed in calcite: both are characterized by lenticular-shaped IFCs and can be excited within the material light cone ($\boldsymbol{q} < n_z$) as long as one in-plane tensor component is negative, while the other two are positive (this is possible in both uniaxial and biaxial crystals). However, leaky modes in calcite are interface waves in semi-infinite uniaxial slabs, while the lenticular modes predicted in $MoOCl_2$ appear as waves solutions for a biaxial thin film and exist outside the light cone of the substrate ($n_z > \boldsymbol{q} > n_{sub}$).

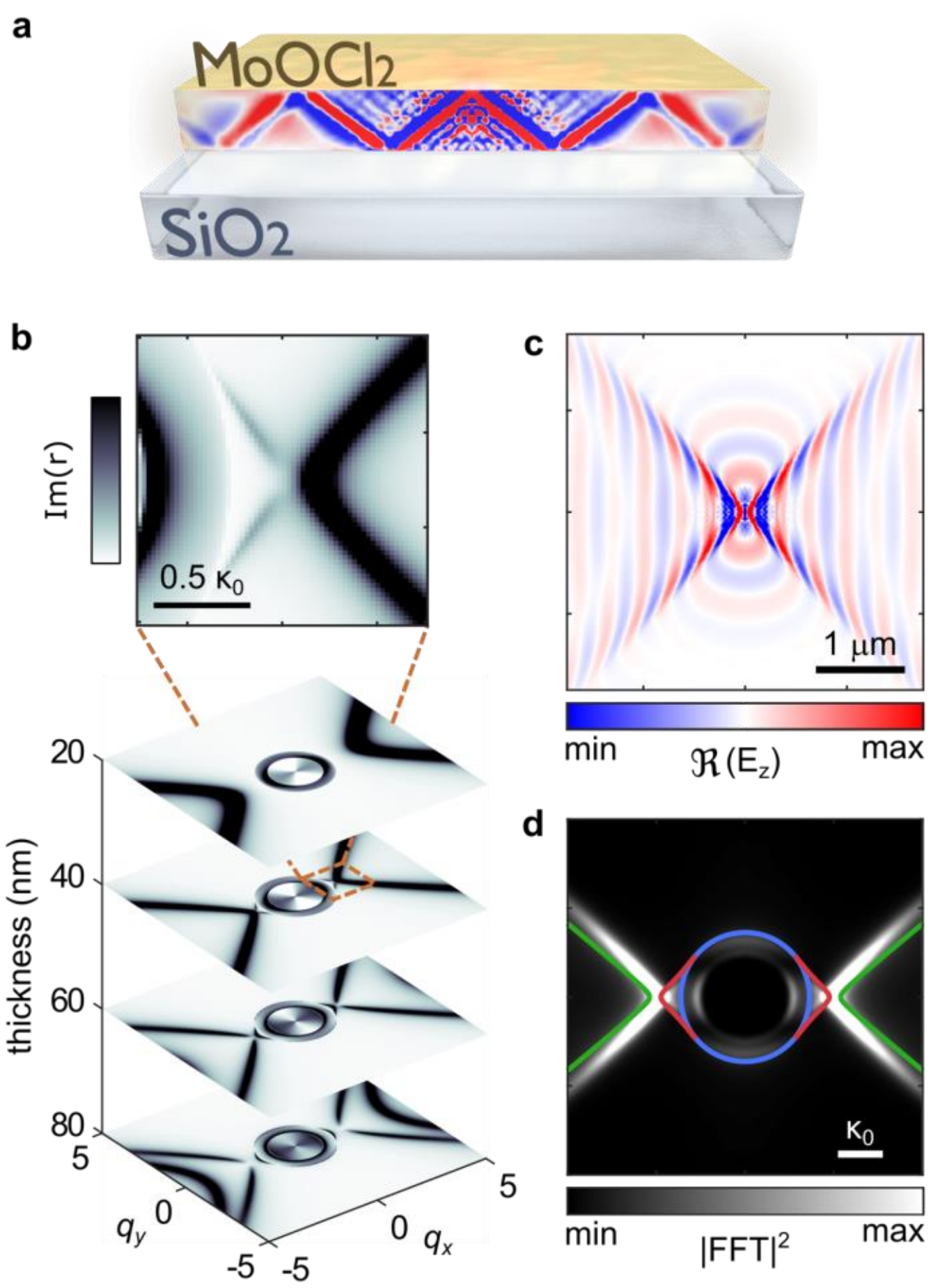

**Figure 3. Plasmon Polaritons propagation in $MoOCl_2$ thin films. a** Schematic representation of PPs in a thin $MoOCl_2$ film on an $SiO_2$ substrate propagating in a ray-light fashion and getting reflected at the top/bottom interfaces. **b** In-plane polariton dispersion calculated with the transfer matrix method for different flake thicknesses. Top inset: zoom view of the intersection between hyperbolic and lenticular modes for a 40 nm flake. $\kappa_0 = 2\pi/\lambda_0$ is the free-space light wavevector. **c** Simulated $\mathrm{Re}(E_z)$ of PPs launched by a near-field source positioned on a 40 nm $MoOCl_2$ slab and **d** corresponding square modulus of the FFT map. Solid lines represent analytical calculations. Hyperbolic PPs are marked in green and lenticular modes in red. Blue line indicates the substrate light-cone at $|\boldsymbol{q}| = n_{sub}$.

In Fig. 3c we report numerical simulations of the $E_z$ field in a 40 nm flake produced by a dipole placed just above the $MoOCl_2$ layer, which shows the excitation of mainly hyperbolic-like PPs. Nevertheless, the (fast-)Fourier transform (FFT) map of the calculated field in Fig. 3d confirms the existence of both the hyperbolic and lenticular PPs predicted by the transfer matrix. To have a further confirmation, we use the recently developed analytical expression for polariton dispersion in biaxial media[50,51] in the small thickness limit ($\kappa_0 d \ll 1$, $\kappa_0 = 2\pi/\lambda_0$) to extract the theoretical PPs dispersions (see Supplementary Information S2). The analytical solution for the PPs, shown as solid lines in Fig. 3d, confirms again the coexistence of a hyperbolic (green) and lenticular (red) dispersion (blue line indicates the $n_{sub} = 1.5$ contour). While lenticular modes have been observed in semi-infinite calcite slabs, their existence as thin-film solutions in $MoOCl_2$ enables the observation of their unique interference with hyperbolic modes.

**Real space imaging of thin-film hyperbolic plasmon polaritons**

In the following, we experimentally confirm the presence of the predicted hyperbolic PPs by real space nano-imaging.

Thin flakes were obtained by mechanical exfoliation of commercially available bulk $MoOCl_2$ on 285 nm $SiO_2$/Si substrates (see Methods). A representative flake seen under the optical microscope is shown in Fig. 4a. Since there are no experimental data in literature about the optical properties of the material, we first look for a confirmation of the anisotropic optical response of the as-exfoliated material. Fig. 4a shows remarkably different flake colours when placing a polarizer in the illumination path, aligned with either the short (left) or long (right) edge of the flake (see Methods). This, in combination with the rectangular shape of the exfoliated flakes (which we observe over many samples, see Supplementary Information S3), suggests that $MoOCl_2$ naturally breaks along the [100] and [010] directions. To provide a direct confirmation of such hypothesis, we perform atomic-resolution atomic force microscopy (AFM, see Methods) of the flake surface (Fig. 4b). Such measurement allows to directly link the rectangular periodic pattern with the detection of the larger Mo and Cl atoms. The extracted in-plane unit cell parameters $a = 4.3$ Å and $b = 7.1$ Å match well with the predicted values[48] (see Supplementary Information S4).

Knowing the crystal axis orientation, to obtain a quantitative evaluation of the optical anisotropy, we take reflectivity spectra of the exfoliated flakes (see Methods). We filter the illumination source with a polarizer rotated parallel, $|x\rangle$, or perpendicular, $|y\rangle$, to the short flake edge corresponding to the [100] crystal axis. The $|x\rangle$-polarized reflectivity (Fig. 4c, yellow curve) is consistent with a metallic response close to the plasma frequency, where the reflectivity decreases as the theoretically predicted $\varepsilon_x$ increases (Fig. S2). The drop in reflectivity below 600 nm explains the yellow colour of the flake, as observed in Fig. 4a (left panel). The spectrum along $|y\rangle$ (Fig. 4c, green curve) shows instead oscillations associated with multiple reflections in the Fabry-Perot cavity formed by the flake, which behaves as a low-loss dielectric slab (with thickness 560 nm for the measurement shown in Fig. 4c) along the [010] direction.

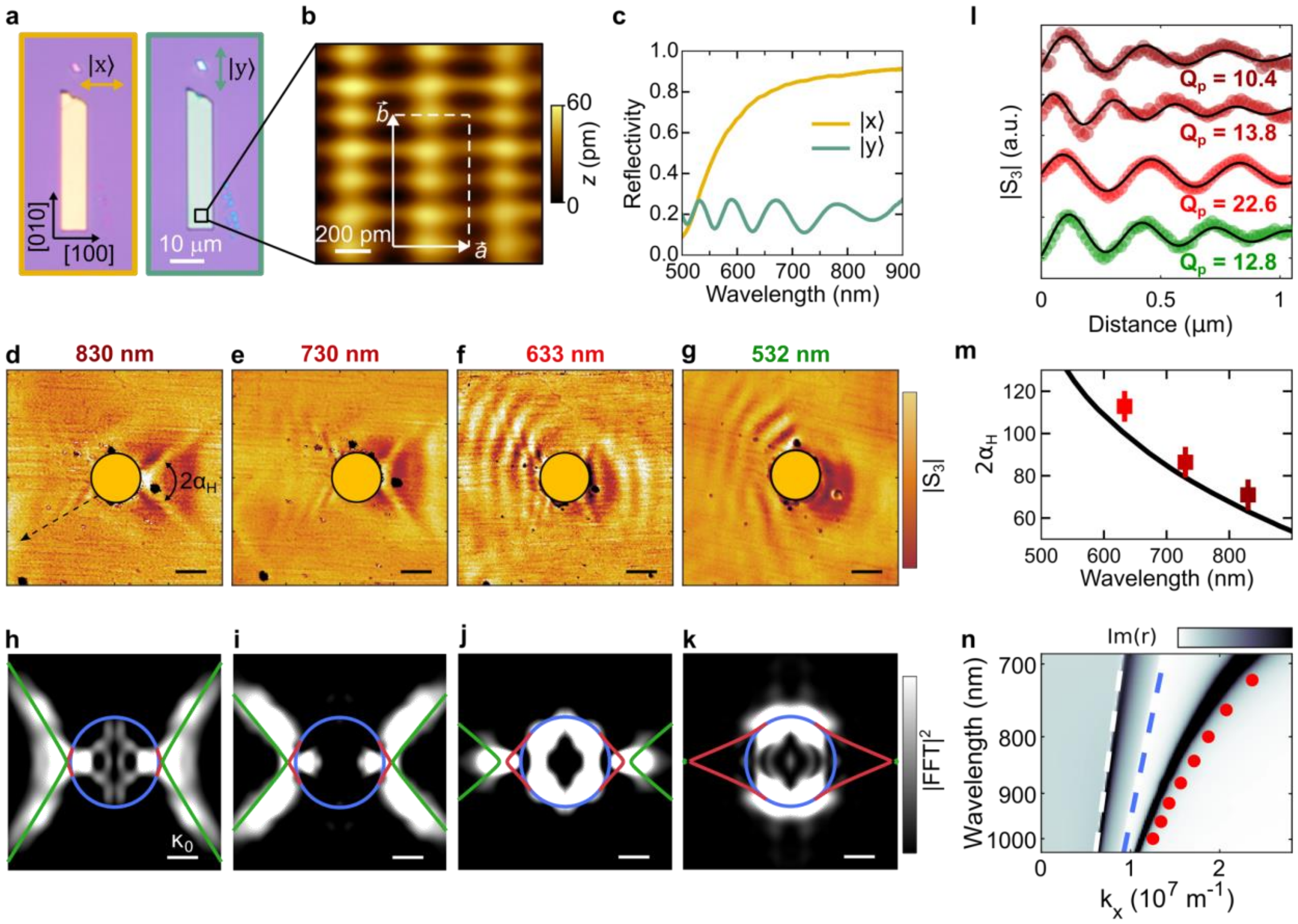

**Figure 4. Characterization of $MoOCl_2$ anisotropy and real-space imaging of hyperbolic PPs at visible and near-IR frequencies. a** Optical images of an as-exfoliated flake with illumination polarized along the $|x\rangle$ (left) and $|y\rangle$ (right) directions. **b** Atomic resolution AFM of the flake surface, where in-plane crystal unit cell vectors are indicated by white arrows. **c** Reflectivity spectra obtained with a polarized source along the $|x\rangle$ (yellow) and $|y\rangle$ (green) directions of a thick flake (~560 nm). **d-g** s-SNOM amplitude maps (third demodulation order) of PPs launched by a gold disk (yellow circle) fabricated on top of the flake (thickness 42 nm) at $\lambda_0$ = 830, 730, 633 and 532 nm (scale bars: 500 nm). **h-k** Corresponding FFT maps with superimposed analytical results of the PPs dispersion in the thin-film approximation (blue: $SiO_2$ light-cone; red: lenticular modes; green: hyperbolic modes). **l** PPs decay extracted along the hyperbolic profiles as shown by dashed line in panel **d.** Solid lines: best-fits of the profiles allowing for the extraction of the decay length. **m** Aperture angle of the hyperbolic PPs extracted from the experiments as shown in panel **d**. Black curve corresponds to theoretical predictions. **n** Experimental dispersion relation of PPs along the metallic direction of a 20 nm flake (red dots) superimposed to the calculated transfer matrix. The white (blue) dashed line corresponds to the light line in air (in the $SiO_2$ substrate).

The experimental verification of the hyperbolic PPs in $MoOCl_2$ thin films requires a near-field excitation and detection scheme. As for the excitation, we fabricate a gold disk on the flake surface (see Methods). Scattering of illuminating light from the disk provides wavevectors beyond the free-space light-cone that can couple to the polaritonic modes of the material. For the detection instead, we use a scattering-type scanning near field optical microscope (s-SNOM) which allows the subwavelength imaging of modes with large in-plane momenta (see Methods)[17,22,23,46,54].

Near-field amplitude maps obtained on a 42 nm-thick flake at $\lambda_0 = 830$, 730, 633 and 532 nm are shown in Fig. 4d-g (the yellow circle indicates the position of the disk). At 830 nm and 730 nm (Fig. 4d, e), we observe hyperbolic wavefronts coming from PPs launched by the gold disk. The asymmetry observed in the oscillations on different sides of the launcher (i.e., left-right sides) is due to the opposite sign of the polariton wavevector with respect to the one of the excitation beams (see Supplementary Information S7)[46]. A small misalignment in the illumination relative to the [100] axis is likely the main contribution to the slight asymmetry between the upper and lower fringes. At 633 nm (Fig 4f), we observe the coexistence of the hyperbolic and lenticular-like modes, as evidenced by the presence of broad, directional wavefronts with a hyperbolic interference (see Supplementary Information S2). The presence of hyperbolic PPs is also highlighted by the ray-like propagation on the right-side of the disk (Fig. 4d-f). At 532 nm (Fig 4g), we only observe the directional propagation from the lenticular IFC as the hyperbolic mode lies at very high wavevectors that are not efficiently excited in our experiment (the momentum imparted by the near-field source depends on its physical dimensions[46], see Supplementary Information S8).

In Fig. 4h-k we show the FFT maps computed from the experimental data in Fig. 4d-g (see Supplementary Information S7) together with the theoretical dispersions calculated in the thin-film limit. We indicate with green, red, and blue lines the hyperbolic, lenticular and substrate (circle with radius $n_{sub} = 1.5$) modes, respectively. We obtain very good agreement with the theoretical dispersion for the long wavelength maps (Fig 4h, i), confirming the presence of hyperbolic PPs in $MoOCl_2$ flakes. For smaller values of $\lambda_0$ (Fig 4j, k), the FFT maps show the presence of the lenticular mode branching from the $n_{sub}$ circle. Here, the agreement with theory diminishes as the thin film approximation tends to overestimate the polariton momentum (see Supplementary Information S2).

From the experimental maps, we extract the PPs propagation length by fitting the oscillations profiles with an exponentially damped sinusoid[54] along tilted lines following the hyperbolic propagation, marked by the dashed arrow in Fig. 4d (see Supplementary Information S9). From the fit, we compute the polariton quality factor $Q_p$ (see Supplementary Information S9) reported in Fig. 4l and estimate a value for the PPs lifetime $\tau_p$ on the order of hundreds of fs (see Supplementary Information S9). Strikingly, while PPs are generally regarded to be lossy, the extracted $Q_p$ are comparable with values reported for PhPs in calcite[54], hBN[55] and $MoO_3$[56].

From the s-SNOM data we also extract the hyperbolic rays' opening angle $2\alpha_H$, as shown in Fig. 4d. We exclude here the 532 nm map, as we do not observe hyperbolic PPs at this frequency (Fig 4g). The extracted data follow the theoretically predicted trend[17,19] determined by $\alpha_{\mathrm{H}} = (\pi/2 - \tan^{-1}\sqrt{-\varepsilon_x/\varepsilon_y})$ (Fig 4m, see also Supplementary Information S2). This is in line with the increase of the angular aperture of the asymptotes $(\pi/2 - \alpha_{\mathrm{H}})$ of the IFCs when increasing $\lambda_0$ (see discussion above and Supplementary Information S2). The experimental values are $< 10\%$ off the predicted values and overlap with the DFT-based[48] theoretical prediction within the associated error (see Supplementary Information S9).

To enrich our observations, we measure the PPs from the edge of a 20 nm-thick crystal along the metallic direction ([100]) for different wavelengths. Since these PPs are tip-launched (see Supplementary Information S10), the observed periodicity is unaffected by the illumination geometry, hence there are no additional uncertainties in retrieving the PPs momentum (differently from the edge-launched PPs from the gold disk, see Supplementary Information S7).[57] When comparing the reconstructed dispersion relation with the results obtained from the transfer matrix method, the experimental values appear shifted towards ~ 10% higher momenta, still preserving the overall trend and overlapping with the theoretical value within the experimental error (Fig. 4n and Supplementary Information S10).

Even if the presented data are in line with our theoretical analysis, the slight discrepancies reported in Fig. 4m, n suggest that the dielectric permittivity of $MoOCl_2$ might differ from the DFT calculations of Ref. [[48]]. Indeed, in both the experimental datasets, the shift could be explained by an actual larger permittivity along the [100] direction in the considered wavelength range. A possible physical justification can be a modification of the actual dielectric response of the material due to electronic correlation effects, not considered in the DFT computations, as suggested in Ref. [[58]].

## Conclusion

$MoOCl_2$ constitutes the first natural material featuring propagating polaritons with hyperbolic contours at near-IR and visible frequencies, which we demonstrated here by real space nano-imaging. Leveraging PPs instead of PhPs is the key element enabling the increase in the operation frequency. While anisotropic PPs have been investigated before, they had only been observed at terahertz frequencies[59] or in the mid-IR upon ultrafast excititation[60]. Our findings extend the control over nanoscale light propagation achieved by PhPs in the mid-IR to the technologically relevant visible range. As $MoOCl_2$ is stable in air without the need of any protection layer, it constitutes and ideal platform to realize applications of hyperbolic materials that until now lacked widespread implementation due to the combination of high losses in metals and limited fabrication resolution in metamaterials.

Besides the presence of hyperbolic polaritons, we believe the unique properties of $MoOCl_2$ can be leveraged for several nanophotonic applications in the visible spectrum. For example, in-plane anisotropic media have recently attracted considerable attention for the realization of ultra compact optical elements for polarization manipulation[61,62]. While several materials show good optical dichroism (i.e., difference in the refractive index imaginary part) at the expense of large losses or low absolute transmission/reflection, $MoOCl_2$ features a wide low-loss frequency range with both large optical dichroism and birefringence[63], making it a good candidate for the realization of ultra compact optical elements with low dissipation. Moreover, thanks to its metallic character with low electron density, $MoOCl_2$ is a promising candidate for optical devices reconfigurable through electrostatic gating, providing an easily accessible alternative to percolated ultra-thin metallic films[64].

To the best of our knowledge, $MoOCl_2$ is the first natural, air-stable vdW material with a plasma frequency in the visible, constituting a novel metallic platform for the emerging field of all-vdW nanophotonics[65,66] which, until now, could only operate with dielectric elements at visible frequencies. Given all the unique properties mentioned above, $MoOCl_2$ could be potentially leveraged to realize high-frequency tuneable miniaturized polarization elements[67] and metasurfaces[68].

## References

1. Basov, D. N., Fogler, M. M. & García De Abajo, F. J. Polaritons in van der Waals materials. *Science (1979)* **354**, (2016).
2. Hirsch, L. R. *et al.* Nanoshell-mediated near-infrared thermal therapy of tumors under magnetic resonance guidance. *Proc Natl Acad Sci U S A* **100**, 13549–13554 (2003).
3. Baffou, G., Cichos, F. & Quidant, R. Applications and challenges of thermoplasmonics. *Nat Mater* **19**, 946–958 (2020).
4. Cortés, E. *et al.* Challenges in Plasmonic Catalysis. *ACS Nano* **14**, 16202–16219 (2020).
5. Herran, M. *et al.* Plasmonic bimetallic two-dimensional supercrystals for H2 generation. *Nature Catalysis 2023 6:12* **6**, 1205–1214 (2023).
6. Xu, Y. *et al.* Optical Refractive Index Sensors with Plasmonic and Photonic Structures: Promising and Inconvenient Truth. *Adv Opt Mater* **7**, 1801433 (2019).
7. Langer, J. *et al.* Present and future of surface-enhanced Raman scattering. *ACS Nano* **14**, 28–117 (2020).
8. Yu, N. *et al.* Light Propagation with Phase Discontinuities: Generalized Laws of Reflection and Refraction Downloaded from.
9. Genevet, P. *et al.* Controlled steering of Cherenkov surface plasmon wakes with a one-dimensional metamaterial. *Nat Nanotechnol* **10**, 804–809 (2015).
10. Wintz, D., Ambrosio, A., Zhu, A. Y., Genevet, P. & Capasso, F. Anisotropic Surface Plasmon Polariton Generation Using Bimodal V-Antenna Based Metastructures. *ACS Photonics* **4**, 22–27 (2017).
11. Lee, J. *et al.* Giant nonlinear response from plasmonic metasurfaces coupled to intersubband transitions. *Nature 2014 511:7507* **511**, 65–69 (2014).
12. Mesch, M., Metzger, B., Hentschel, M. & Giessen, H. Nonlinear Plasmonic Sensing. *Nano Lett* **16**, 3155–3159 (2016).
13. Wang, H. *et al.* Planar hyperbolic polaritons in 2D van der Waals materials. *Nature Communications 2024 15:1* **15**, 1–14 (2024).
14. Poddubny, A., Iorsh, I., Belov, P. & Kivshar, Y. Hyperbolic metamaterials. *Nature Photonics 2013 7:12* **7**, 948–957 (2013).
15. Galiffi, E. *et al.* Extreme light confinement and control in low-symmetry phonon-polaritonic crystals. *Nature Reviews Materials 2023 9:1* **9**, 9–28 (2023).
16. Tamagnone, M. *et al.* Ultra-confined mid-infrared resonant phonon polaritons in van der Waals nanostructures. **4**, eaat7189 (2018).
17. Dai, S. *et al.* Tunable phonon polaritons in atomically thin van der Waals crystals of boron nitride. *Science (1979)* **343**, 1125–1129 (2014).
18. Ambrosio, A. *et al.* Selective excitation and imaging of ultraslow phonon polaritons in thin hexagonal boron nitride crystals. *Light Sci Appl* **7**, (2018).

19. Li, P. *et al.* Hyperbolic phonon-polaritons in boron nitride for near-field optical imaging and focusing. *Nature Communications 2015 6:1* **6**, 1–9 (2015).

20. Ambrosio, A. *et al.* Mechanical Detection and Imaging of Hyperbolic Phonon Polaritons in Hexagonal Boron Nitride. *ACS Nano* **11**, 8741–8746 (2017).

21. Chaudhary, K. *et al. Engineering Phonon Polaritons in van Der Waals Heterostructures to Enhance In-Plane Optical Anisotropy*. http://advances.sciencemag.org/ (2019).

22. Zheng, Z. *et al.* A mid-infrared biaxial hyperbolic van der Waals crystal. *Sci Adv* **5**, (2019).

23. Ma, W. *et al.* In-plane anisotropic and ultra-low-loss polaritons in a natural van der Waals crystal. *Nature* **562**, 557–562 (2018).

24. Passler, N. C. *et al.* Hyperbolic shear polaritons in low-symmetry crystals. *Nature 2022 602:7898* **602**, 595–600 (2022).

25. Matson, J. *et al.* Controlling the propagation asymmetry of hyperbolic shear polaritons in beta-gallium oxide. *Nature Communications 2023 14:1* **14**, 1–8 (2023).

26. Zhang, Q. *et al. Unidirectionally Excited Phonon Polaritons in High-Symmetry Orthorhombic Crystals*.

27. Hu, G. *et al.* Topological polaritons and photonic magic angles in twisted α-MoO3 bilayers. *Nature* **582**, 209–213 (2020).

28. Álvarez-Pérez, G. *et al.* Negative reflection of nanoscale-confined polaritons in a low-loss natural medium. *Sci Adv* **8**, 8486 (2022).

29. Sternbach, A. J. *et al.* Negative refraction in hyperbolic hetero-bicrystals. *Science (1979)* **379**, 555–557 (2023).

30. Wu, X. & Fu, C. Near-field radiative heat transfer between uniaxial hyperbolic media: Role of volume and surface phonon polaritons. *J Quant Spectrosc Radiat Transf* **258**, 107337 (2021).

31. Bylinkin, A. *et al.* Real-space observation of vibrational strong coupling between propagating phonon polaritons and organic molecules. *Nat Photonics* **15**, 197–202 (2021).

32. Liu, Z., Lee, H., Xiong, Y., Sun, C. & Zhang, X. Far-field optical hyperlens magnifying sub-diffraction-limited objects. *Science (1979)* **315**, 1686 (2007).

33. Jacob, Z. *et al.* Active hyperbolic metamaterials: enhanced spontaneous emission and light extraction. *Optica, Vol. 2, Issue 1, pp. 62-65* **2**, 62–65 (2015).

34. Lee, Y. U. *et al.* Hyperbolic material enhanced scattering nanoscopy for label-free super-resolution imaging. *Nature Communications 2022 13:1* **13**, 1–8 (2022).

35. Sun, J. & Litchinitser, N. M. Toward Practical, Subwavelength, Visible-Light Photolithography with Hyperlens. *ACS Nano* **12**, 542–548 (2018).

36. High, A. A. *et al.* Visible-frequency hyperbolic metasurface. *Nature 2015 522:7555* **522**, 192–196 (2015).

37. Lee, D. *et al.* Realization of Wafer-Scale Hyperlens Device for Sub-diffractional Biomolecular Imaging. *ACS Photonics* **5**, 2549–2554 (2018).

38. Kang, E. S. H. *et al.* Organic Anisotropic Excitonic Optical Nanoantennas. *Advanced Science* **9**, 2201907 (2022).

39. Lee, Y. U., Yim, K., Bopp, S. E., Zhao, J. & Liu, Z. Low-Loss Organic Hyperbolic Materials in the Visible Spectral Range: A Joint Experimental and First-Principles Study. *Advanced Materials* **32**, 2002387 (2020).

40. Esslinger, M. *et al.* Tetradymites as Natural Hyperbolic Materials for the Near-Infrared to Visible. *ACS Photonics* **1**, 1285–1289 (2014).

41. Córdova-Castro, R. M. *et al.* Anisotropic Plasmonic CuS Nanocrystals as a Natural Electronic Material with Hyperbolic Optical Dispersion. *ACS Nano* **13**, 6550–6560 (2019).

42. Shao, Y. *et al.* Infrared plasmons propagate through a hyperbolic nodal metal. *Sci Adv* **8**, 6169 (2022).

43. Korzeb, K. *et al.* Compendium of natural hyperbolic materials. *Optics Express, Vol. 23, Issue 20, pp. 25406-25424* **23**, 25406–25424 (2015).

44. Gjerding, M. N., Petersen, R., Pedersen, T. G., Mortensen, N. A. & Thygesen, K. S. Layered van der Waals crystals with hyperbolic light dispersion. *Nature Communications 2017 8:1* **8**, 1–8 (2017).

45. Ruta, F. *et al.* Hyperbolic exciton polaritons in a van der Waals magnet. doi:10.21203/rs.3.rs-3239594/v1.

46. Ma, W. *et al.* Ghost hyperbolic surface polaritons in bulk anisotropic crystals. *Nature* **596**, 362–366 (2021).

47. Gao, H., Ding, C., Sun, L., Ma, X. & Zhao, M. Robust broadband directional plasmons in a MoOCl2 monolayer. *Phys Rev B* **104**, 205424 (2021).

48. Zhao, J. *et al.* Highly anisotropic two-dimensional metal in monolayer MoOCl2. *Phys Rev B* **102**, (2020).

49. Wang, Z. *et al.* Fermi liquid behavior and colossal magnetoresistance in layered MoOC l2. *Phys Rev Mater* **4**, (2020).

50. Álvarez-Pérez, G., Voronin, K. V., Volkov, V. S., Alonso-González, P. & Nikitin, A. Y. Analytical approximations for the dispersion of electromagnetic modes in slabs of biaxial crystals. *Phys Rev B* **100**, 235408 (2019).

51. Narimanov, E. E. Dyakonov waves in biaxial anisotropic crystals. *Phys Rev A (Coll Park)* **98**, 013818 (2018).

52. Ma, W. *et al.* Ghost hyperbolic surface polaritons in bulk anisotropic crystals. *Nature* **596**, 362–366 (2021).

53. Simovski, C. & Tretyakov, S. *An Introduction to Metamaterials and Nanophotonics*. (Cambridge University Press, Cambridge, 2020). doi:DOI: 10.1017/9781108610735.

54. Ni, X. *et al.* Observation of directional leaky polaritons at anisotropic crystal interfaces. *Nat Commun* **14**, (2023).

55. Giles, A. J. *et al.* Ultralow-loss polaritons in isotopically pure boron nitride. *Nature Materials 2017 17:2* **17**, 134–139 (2017).

56. Menabde, S. G. *et al.* Low-Loss Anisotropic Image Polaritons in van der Waals Crystal α-MoO3. *Adv Opt Mater* **10**, 2201492 (2022).

57. Kaltenecker, K. J. *et al.* Mono-crystalline gold platelets: A high-quality platform for surface plasmon polaritons. *Nanophotonics* **9**, 509–522 (2020).

58. Ruta, F. L. *et al.* Good plasmons in a bad metal. *arXiv:2406.05703* (2024).

59. Chen, S. *et al.* Real-space observation of ultraconfined in-plane anisotropic acoustic terahertz plasmon polaritons. *Nature Materials 2023 22:7* **22**, 860–866 (2023).

60. Fu, R. *et al.* Manipulating hyperbolic transient plasmons in a layered semiconductor. *Nature Communications 2024 15:1* **15**, 1–8 (2024).

61. Niu, S. *et al.* Giant optical anisotropy in a quasi-one-dimensional crystal. *Nature Photonics 2018 12:7* **12**, 392–396 (2018).

62. Guo, Q. *et al.* Colossal in-plane optical anisotropy in a two-dimensional van der Waals crystal. *Nature Photonics 2024* 1–6 (2024) doi:10.1038/s41566-024-01501-3.

63. Zhou, Y. *et al.* A solution-processable natural crystal with giant optical anisotropy for efficient manipulation of light polarization. *Nature Photonics 2024 18:9* **18**, 922–927 (2024).

64. Maniyara, R. A. *et al.* Tunable plasmons in ultrathin metal films. *Nature Photonics 2019 13:5* **13**, 328–333 (2019).

65. Meng, Y. *et al.* Photonic van der Waals integration from 2D materials to 3D nanomembranes. *Nature Reviews Materials 2023 8:8* **8**, 498–517 (2023).

66. Weber, T. *et al.* Intrinsic strong light-matter coupling with self-hybridized bound states in the continuum in van der Waals metasurfaces. *Nature Materials 2023 22:8* **22**, 970–976 (2023).

67. Lynch, J. *et al.* Gate-tunable optical anisotropy in wafer-scale, aligned carbon nanotube films. *Nature Photonics 2024* 1–9 (2024) doi:10.1038/s41566-024-01504-0.

68. Jung, C., Lee, E. & Rho, J. The rise of electrically tunable metasurfaces. *Sci Adv* **10**, eado8964 (2024).

69. Paarmann, A. & Passler, N. C. Generalized 4 × 4 matrix formalism for light propagation in anisotropic stratified media: study of surface phonon polaritons in polar dielectric heterostructures. *JOSA B, Vol. 34, Issue 10, pp. 2128-2139* **34**, 2128–2139 (2017).

**Methods** (including separate data and code availability statements)

***Electromagnetic simulations*** **-** Electromagnetic simulations were performed with a commercial solver (CST Studio) in frequency domain. The dielectric function of $MoOCl_2$ was taken from theoretical calculations in Ref. [48] and modelled as an anisotropic media in CST. We place a dipole-like near-field source immersed in the middle of a slab of $MoOCl_2$ to excite all the PPs modes with high wavevectors beyond the free-space light momentum $\kappa_0$[26]. The near-field source was obtained by inserting a discrete port in a vacuum gap inside a perfect conducting cylinder with rounded caps (see Supplementary Information S8). As CST does not support open boundaries for anisotropic media, a vacuum gap was inserted between the end of the $MoOCl_2$ slab and the open boundary conditions. As port modes touching anisotropic media are also not supported, the near-field source in the simulations shown in Fig. 2b, c was placed in a small vacuum enclosure in the middle of the $MoOCl_2$ slab.

***Transfer matrix technique*** **-** The calculations of the imaginary part of the reflection coefficients were performed using a 4×4 transfer matrix formalism[69]d components, for a stack of homogenous media with arbitrary anisotropic dielectric tensor. The two-dimensional maps in Fig. 3b were obtained by rotating the $MoOCl_2$ dielectric function around the z-axis and computing the reflection coefficients at each angle.

***Mechanical Exfoliation*** – two different substrates were used: fused silicon oxide ($SiO_2$) and boron-doped Si wafer with 285 nm of thermally grown $SiO_2$ on top ($SiO_2$/Si), both purchased (Wafer University). Before the exfoliation, the substrates were soaked for 5 minutes in acetone to remove organic residues and rinsed with isopropyl alcohol (IPA). The thin flakes were exfoliated by the standard mechanical exfoliation method starting from a $MoOCl_2$ bulk crystal (HQ Graphene), using commercially available processing tape purchased from Ultron Systems, Inc. The exfoliated samples were cleaned with a 30-minute bath in N-Methyl-2-Pyrollidone (NMP) at 80°C to remove the residues from the exfoliation, then bathed in acetone for 5 minutes and rinsed with IPA.

***Atomic resolution Atomic Force Microscopy*** – The AFM experiments related to Fig. 4b were performed in air, at room temperature, using a Multimode 8 (Bruker) AFM microscope. ARROW-UHF (NanoAndMore) cantilevers (spring constant ≈ 40 $Nm^{-1}$) were used as received. The images were obtained in contact mode, with a set-point of about 5 nm. Once the final scan size of about 10 nm x 10 nm was reached, the scan rate was set to about 20 Hz to minimize the thermal drift and enhance the atomic resolution contrast. The AFM images were analysed with the Gwyddion software (version 2.65). First, the data were levelled by mean plane subtraction and the row alignment tool was applied. The data were then FFT-filtered, and the residual drift was corrected with the Drift Correction tool. The measurements of the lattice constants were extracted with the Lattice tool in the software. Unfiltered data are shown in Fig. S14 in Supplementary Information S4.

***Far-field optical characterization*** – the images shown in Fig. 4a and Fig. S12, S13 in the Supplementary Information were taken in an optical microscope. For the polarized measurements, the incoming light was polarized with a broadband polarizer. No polarizer was put in the collection path. For the cross-polarization measurements (see Fig. S12 in the Supplementary Information), a polarizer was added in collection. The reflection measurements shown in Fig. 4c were taken by placing samples in an inverted microscope (Nikon Eclipse Ti2) coupled to an optical spectrometer (Andor Kymera 328) equipped with a wide range grating (150 lines/mm) and an electrically cooled CCD (Andor Newton). A supercontinuum laser (SuperK Extreme by NKT Photonics) was employed as illumination source. The incoming light was polarized with a wire-grid polarizer (ThorLabs). The samples were exfoliated on $SiO_2$ to minimize the optical contribution of the substrate. All the measurements were normalized to the reflection of a silver mirror measured in the same illumination condition.

***Fabrication of the gold antennas*** – the antennas used in the near-field experiments were fabricated by means of electron beam lithography on flakes exfoliated on $SiO_2$/Si substrates. The sample was spin-coated with an e-beam positive resist (PMMA 950k) for 1 minute at 4000 rpm and baked on a hot plate for 3 minutes at 160° C to remove the solvents. After the lithography, the resist was developed by soaking the sample for 1 minute in a solution 1:3 of (Methyl Isobutyl Ketone):IPA at room temperature. 5 nm of chromium (used as adhesive layer), followed by 45 nm of high purity gold were then deposited in high vacuum in a thermal evaporator. The sample was left in acetone overnight to lift-off the resist mask and then cleaned with a 30-minute bath in NMP at 80° C. Finally, the sample was bathed in acetone and rinsed with IPA. For more details, see Supplementary Information S6.

***Scattering scanning near-field optical microscopy*** – Near-field maps of Fig. 4d-g were obtained with a commercial s-SNOM set-up (Neaspec). Briefly, a laser beam is focused by an off-axis parabolic mirror under the apex of a metal-coated (Pt/Ir) AFM tip (with polarization along the tip elongated axis) operating in tapping mode (tapping amplitude $\approx 50$ nm) over the sample (gold disk on $MoOCl_2$), which is raster scanned below the tip, and the backscattered signal is collected by the same parabolic mirror and sent to the detector. The tip acts as a moving near-field source itself, providing the required momenta to launch PPs. The modes launched by the tip are scattered by the gold disk or vice versa. A modulated far-field signal at the detector plane results from the interference of the light scattered from the tip and the disk. This modulation is a function of the tip-disk distance and results in the imaging of the propagating modes. To isolate the near field component, the signal is demodulated at higher harmonics ($n\Omega, n = 3$) of the cantilever oscillation frequency ($\Omega \approx 280$ kHz). Before focusing, half of the light is redirected towards a pseudo-heterodyne interferometer by a beam splitter. The light scattered by the tip is then recombined with the interferometer reference to retrieve amplitude and phase resolved maps of the electromagnetic field at the sample surface. To optimize the PPs scattering, we rotate the $MoOCl_2$ flake so that the [100] axis is aligned with the laser incident direction (horizontal direction in Fig. 4d-g). The light sources used in the experiments are a Titanium-Sapphire laser covering the 700-1000 nm range (SolsTiS, M2), a 633 nm diode laser (Match Box, Integrated Optics) and a 532 nm diode laser (LCX-532, Oxxius). All the measurements were done in air and at room temperature.

**Data Availability**

Data are available upon reasonable request to the corresponding authors.

**Acknowledgements**

The authors would like to thank Lin Nan for the help with the 3D rendering of the graphics. The authors would also like to acknowledge Dmitri Basov and Francesco Ruta for the insightful discussion about the interpretation of the experimental data.

This work has been financially supported by the European Research Council (ERC) under the European Union Horizon 2020 research and innovation programme "METAmorphoses", grant agreement no. 817794.

**Competing interest declaration**

The authors declare no competing interests.

**Author contributions**

A. M. conceived the idea. G. V. and N. M. fabricated the samples. N. M. and G. V. performed the far-field measurements. S. C. carried out the atomic force microscopy experiments. G. V. and A. M. performed the near-field measurements. G. V. and A. M. carried out the numerical simulations and theoretical analysis. G.V., A. M., N. M. and A.A. wrote the paper with contributions from all other authors. All the authors discussed the implementation of the research, the experimental data and the related analysis. A.A. coordinated the research activities.

# Supplementary Information

*Giacomo Venturi*, Andrea Mancini*, Nicola Melchioni*, Stefano Chiodini, Antonio Ambrosio*

G. Venturi, A. Mancini, N. Melchioni, S. Chiodini, A. Ambrosio

Centre for Nano Science and Technology, Fondazione Istituto Italiano di Tecnologia, Via Rubattino 81, Milano 20134, Italy

G. Venturi

Physics Department, Politecnico Milano, Piazza Leonardo Da Vinci 32, Milano 20133, Italy

*These authors contributed equally to this work

## S1 – Electronic and optical properties of bulk $MoOCl_2$

### Band Structure

The band structure of $MoOCl_2$, reported in Ref. [1], features 4 low-energy bands which are mostly related to the Mo 4$d$ orbitals. As explained in the main text, $MoOCl_2$ is a Van der Waals layered crystal. Thanks to this, the dispersion of the bands in the out-of-plane direction is very slow and we can limit the analysis of the in-plane structure to the bands computed for monolayer $MoOCl_2$ to simplify the discussion.

The in-plane bands (reported in Ref. [2]) show a large dispersion along the $k_x$ direction (along $\Gamma - \mathrm{X}$), which corresponds in real space to the direction parallel to the Mo-O chains. This, combined with the very little dispersion along the $k_y$ directions, allows to consider $MoOCl_2$ as a system of weakly coupled 1D Mo-O chains[1,2].

The anisotropic optical response in the visible stems from this particular band structure. The first single electron transition happens from the highest valence band to the lowest unoccupied conduction band minimum (corresponding to a DOS maximum above the Fermi energy). In the case of $MoOCl_2$, this transition happens between the valence band at around $-2$ eV and the maximum DOS at 1 eV. Both the starting and ending bands are mostly flat, thus resulting in a Lorentzian-like joint density of states and an expected resonant absorption at around 3 eV.

The plasma frequency tensor is linked to the electron effective mass, and can be computed from the band dispersion with the formula[2]:

$$\left(\omega^{p}_{\alpha,\beta}\right)^2 = -\frac{4\pi e^2}{V}\sum_{n,\boldsymbol{k}} 2f'_{n,\boldsymbol{k}}\left(\hat{\boldsymbol{e}}_\alpha \cdot \frac{\partial E_{n,\boldsymbol{k}}}{\partial \boldsymbol{k}}\right)\left(\hat{\boldsymbol{e}}_\beta \cdot \frac{\partial E_{n,\boldsymbol{k}}}{\partial \boldsymbol{k}}\right)$$

where $V$ is the volume of the unitary cell, $f_{n,\boldsymbol{k}}$ is the equilibrium distribution function, $\hat{\boldsymbol{e}}_\alpha$ and $\hat{\boldsymbol{e}}_\beta$ are the unit vector along the $\alpha$ and $\beta$ directions respectively and $n$ is the band index. By this formula, the plasma frequency is linked to the first derivative of the energy dispersion. Considering the band structure of $MoOCl_2$, the dispersion along the $k_x$ direction is stronger with respect to the other directions, thus:

$$\omega^{p}_{xx} \gg \omega^{p}_{yy}, \omega^{p}_{zz}$$

All these considerations reflect in the computed dielectric functions, reported in Fig. S1 for the wavelength range between 300 nm and 10 μm. First, the plasma frequencies along the $y$ and $z$ directions are in the mid infrared region of the spectrum, much lower than the plasma frequency along $x$ which is in the visible. Moreover, the large Lorentzian-like shapes visible in the in-plane response and centred at ~2.9 eV are associated to the first electronic transitions, as explained above.

### Dielectric permittivity model

The dielectric permittivity of monolayer and bulk $MoOCl_2$ are reported in Ref. [1], as computed with density functional theory (DFT). We modelled the dielectric function of bulk $MoOCl_2$ with a Drude-Lorentz (DL) model:

$$\varepsilon_{xx} = \varepsilon^{x}_{\infty} - \frac{\left(\omega^{p}_{xx}\right)^2}{\omega(\omega + i\gamma^{D}_{x})} + \frac{f^{x}_{1}\left(\omega^{p}_{xx}\right)^2}{\omega^2_{x,1} - \omega^2 - i\gamma_{x,1}\omega}$$

$$\varepsilon_{yy} = \varepsilon_\infty^y - \frac{(\omega_{yy}^p)^2}{\omega(\omega + i\gamma_y^D)} + \sum_{j=1}^{2} \frac{f_j^y(\omega_{yy}^p)^2}{\omega_{y,j}^2 - \omega^2 - i\gamma_{y,j}\omega}$$

$$\varepsilon_{zz} = \varepsilon_\infty^z - \frac{(\omega_{zz}^p)^2}{\omega(\omega + i\gamma_D^z)}$$

where $\varepsilon_\infty^k$ (with $k = \{x, y, z\}$) is the dielectric permittivity at infinite frequency for component $k$, $\omega_{kk}^p$ and $\gamma_D^k$ are the plasma frequency and the damping rate of the Drude response, $f_j^k$ are weight parameters for each interband transition respectively, $\omega_{k,j}$ is the central frequency of each transition while $\gamma_{k,j}$ is the damping rate of the $j$-th Lorentzian response for component $k$. We extracted the parameters of the model by numerically fitting the DFT data from Ref. [2]. The values extracted from the fit are reported in Tab. S1. The DFT dielectric function is well represented by the DL model, as visible in Fig. S2.

| | $\varepsilon_\infty$ | $\omega_P$ | $\gamma_D$ | $f_1$ | $\omega_1$ | $\gamma_1$ | $f_2$ | $\omega_2$ | $\gamma_2$ |
|---|---|---|---|---|---|---|---|---|---|
| $\varepsilon_{xx}$ | 2.74 | 6.419 | 0.023 | 0.42 | 3.28 | 0.46 | | | |
| $\varepsilon_{yy}$ | 3.69 | 1.361 | 0.016 | 1.44 | 0.79 | 0.29 | 14 | 2.84 | 0.13 |
| $\varepsilon_{zz}$ | 3.96 | 0.562 | 0.025 | | | | | | |

**Table S1. Parameter values for the Drude-Lorentz model.** All the values are reported in eV, except for $\varepsilon_\infty$, $f_0$ and $f_j$ which are adimensional numbers.

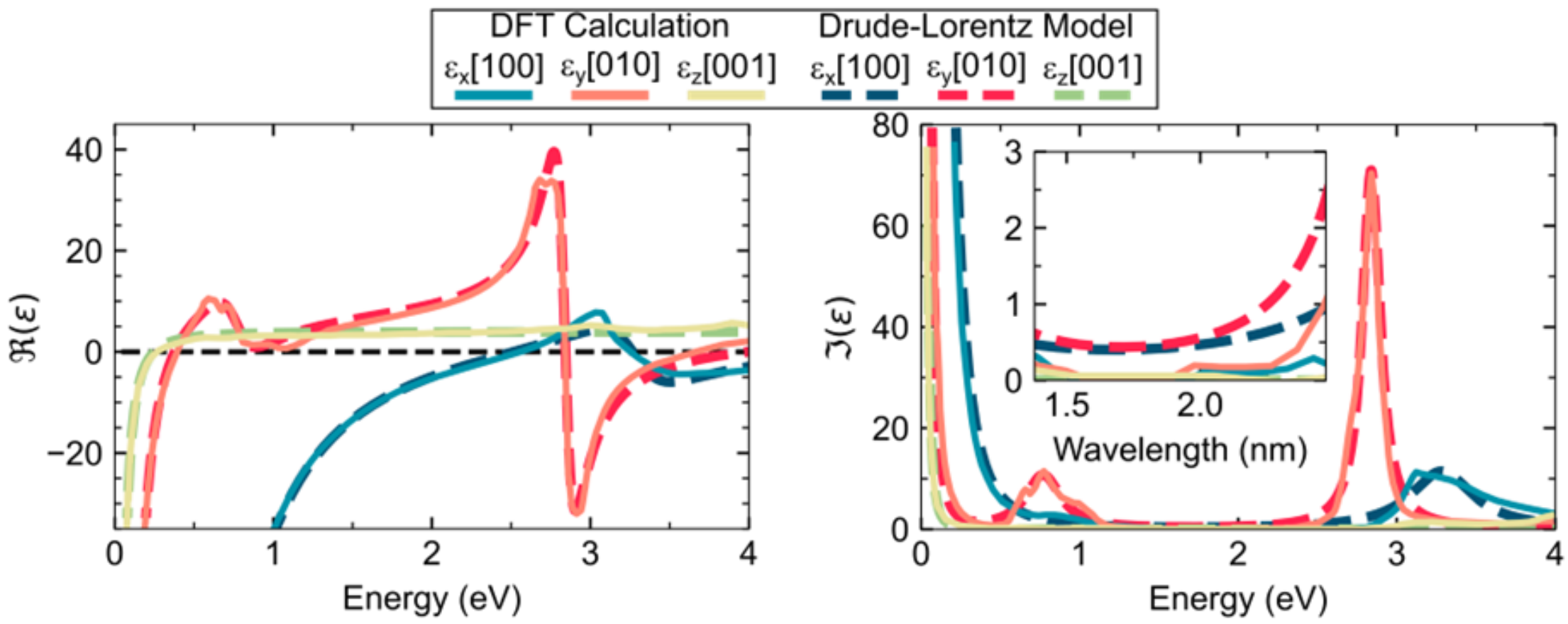

**Figure S1. Dielectric permittivity of $MoOCl_2$.** Comparison between the DFT calculations reported in Ref. 1 (solid lines) and the Drude-Lorentz model developed in the present work (dashed lines).

In the region of interest for the presented experiment (between 500 and 900 nm, see dispersion reported in nm in Fig. S2), the real part of the dielectric function is negative along the $x$ direction due to the Drude response with a large plasma frequency ($\omega_P^x = 6.42$ eV). On the other hand, both the plasma frequencies of the $y$ and $z$ components are below this energy region, and the dielectric functions are positive. On the other hand, the imaginary part of the dielectric functions (i.e., the material optical losses) are small, especially along the metallic $x$ direction, as visible from the inset in the right panel of Fig. S1.

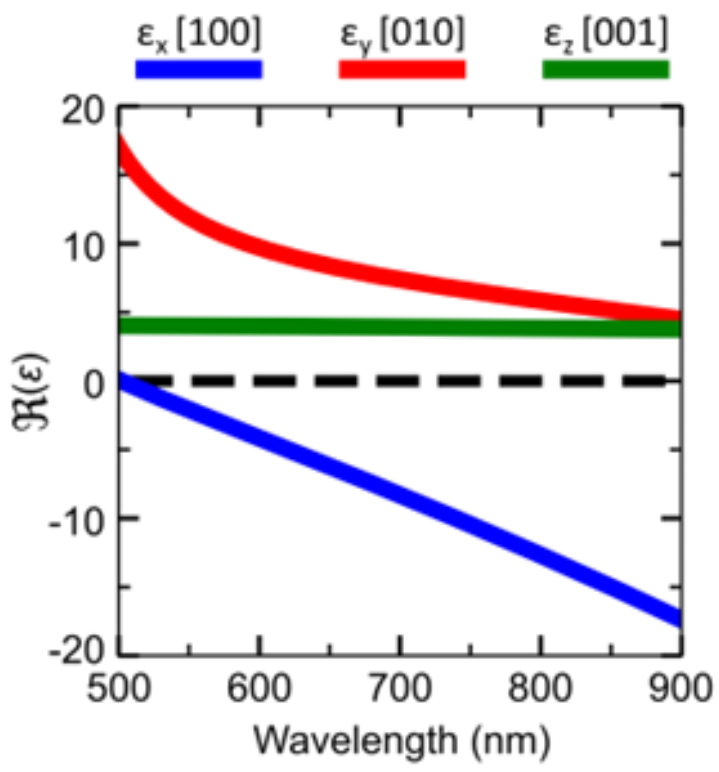

**Figure S2. Real part of the dielectric permittivity of $MoOCl_2$.** Close up of the modelled dielectric function in the region between 500 nm and 900 nm.

**Comparison with other natural hyperbolic materials**

Two figures of merit are commonly used to describe the hyperbolicity of materials: Q factor and dielectric anisotropy (DA)[3]. The DA is defined as $\Delta\varepsilon = \Re(\varepsilon_i) - \Re(\varepsilon_j)$, where $i$ and $j$ indicates the dielectric and metallic directions, respectively. The Q factor gives information about the electromagnetic losses in the metallic direction of the material and is defined as:

$$Q_j = -\frac{\Re(\varepsilon_j)}{\Im(\varepsilon_j)}$$

where $j$ is the metallic direction (for which $\Re(\varepsilon_j) < 0$).

In recent years, different natural materials showing hyperbolic behaviour in the visible range have been proposed. A comparison of the figures of merit for such materials is shown in Fig. S3a, where the different materials are classified by the maximum Q factor and the relative $\Delta\varepsilon$, computed at the corresponding wavelength $\lambda_0$. The radius of each dot is proportional to the width of the wavelength region for which the material is anisotropic. GaTe[3] is a biaxial crystal with both in-plane and out-of-plane anisotropy in the high-energy region of the visible spectrum (from 250 nm to more than 500 nm). The tetradymites[4] $Bi_2Se_3$ and $Bi_2Te_3$ and the metal diborides[5] $MgB_2$, $CaB_2$, $SrB_2$ and $CrB_2$ are instead uniaxial crystals which show out-of-plane anisotropy. Although hyperbolic in the visible, all these crystals suffer from very large losses, with Q factors lower than 1. A better result was achieved with CrSBr[6], a van der Waals magnet which shows anisotropy due to excitons when cooled at cryogenic temperature (20 K). Even if the DA is very large ($\sim 160$), the maximum Q factor is of the order of 1, and only in a very small wavelength region from 855 nm to 911 nm. This, in combination with the need for cryogenic temperatures, strongly limits the applicability of CrSBr as a hyperbolic optical material. More recently, organic hyperbolic materials were proposed. In particular, the quinoidal oligothiophene derivative QQT(CN)4[7] shows hyperbolicity between 620 nm and 970 nm with a very low Q factor. Another organic material, regioregular poly(3-hexylthiophene-2,5-diyl) (rr-P3HT)[8], features a high Q factor of 18, but the DA is very low, and the hyperbolic region is limited in between 400 nm and 550 nm.

On the other hand, $MoOCl_2$ shows a remarkably high Q factor much larger than 1 (with a maximum larger than 32) in the whole spectral region studied in our experiment. In the same region, the dielectric losses along the $y$ direction are also very low ($\Im(\varepsilon_y) \sim 0.5$, as shown in Fig. S3b). Interestingly, the Q factor of hyperbolic PPs in $MoOCl_2$ is comparable to the Q factor computed for common hyperbolic materials in which the anisotropy arises from the presence of phonon polaritons (Fig. S3d shows the Q factor for hBN[9], calcite[10], $\alpha V_2O_5$[11] and $\alpha MoO_3$[12]). In addition to the high Q factor, the DA of $MoOCl_2$ varies from 14 to 21 (see Fig. S3c), and the predicted in-plane hyperbolic region extends from 500 nm to 3000 nm (Fig S1). All these properties make $MoOCl_2$ stand out as a low-loss highly anisotropic material, suitable for applications in visible optics and polaritonics, as discussed in the main text.

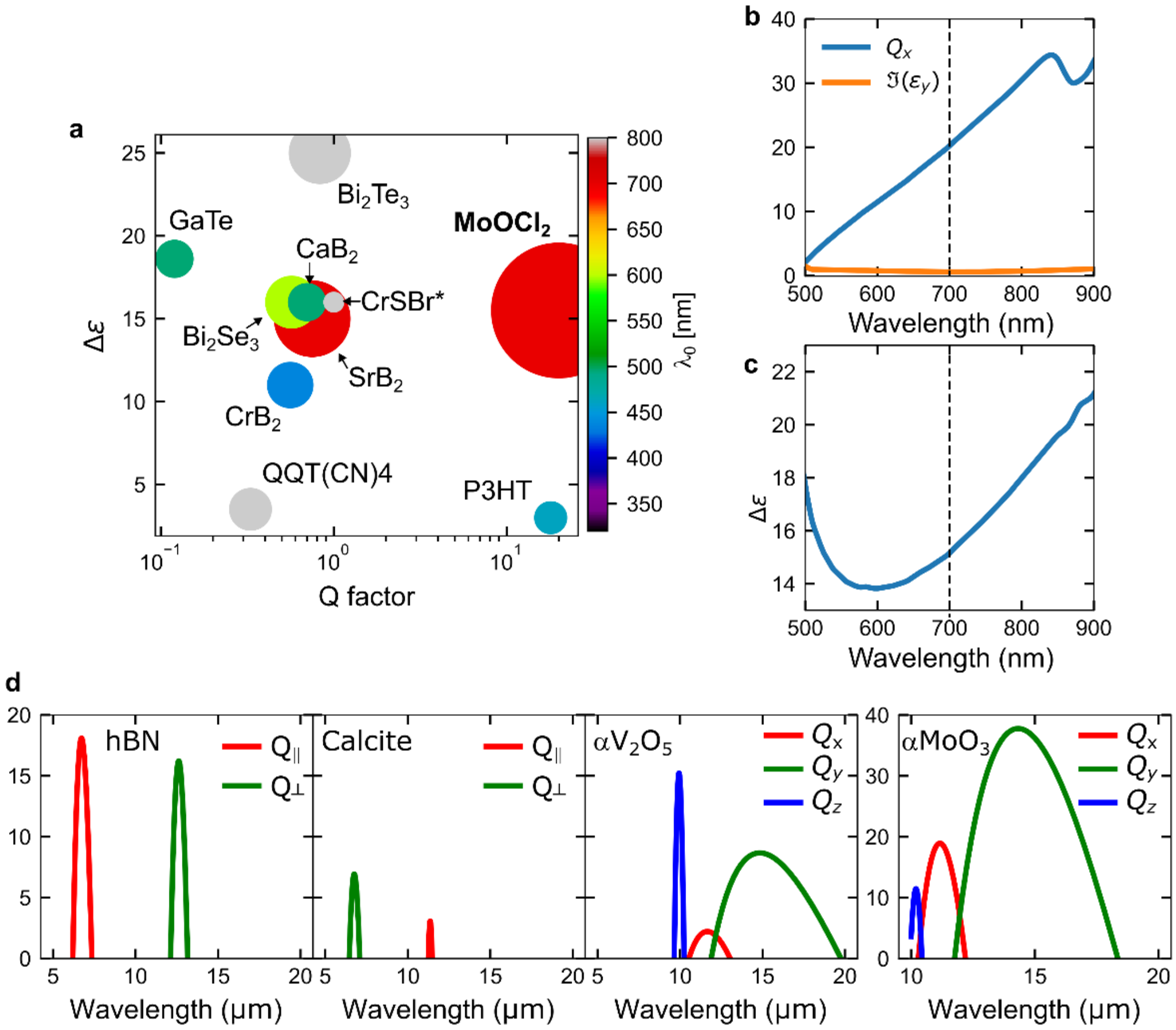

**Figure S3. Q factor and dielectric anisotropy of different hyperbolic materials. a** Comparison between the Q factor and the DA of different hyperbolic materials reported in literature. $\lambda_0$ is the wavelength at which data were taken for each material. The centre of the region of interest (700 nm, vertical dashed lines in **b** and **c**) was taken for $MoOCl_2$. The radius of each dot is proportional to the width of the hyperbolicity region. *The DA of CrSBr shown in the plot is divided by 10. **b** Q factor along $x$ (blue line) and imaginary part of the dielectric function along $y$ of bulk $MoOCl_2$. **c** In-plane dielectric anisotropy between the $x$ and $y$ directions of $MoOCl_2$. **d** Q factor for different anisotropic materials arising from phonon polaritons.

**S2 – Analytical calculation of the isofrequency contours**

**Bulk $MoOCl_2$**

The propagation of light in a bulk uniaxial (and non-magnetic) crystal is described by the solutions of Maxwell's equations for an infinite medium with dielectric tensor:

$$\varepsilon = \begin{bmatrix} \varepsilon_x & 0 & 0 \\ 0 & \varepsilon_y & 0 \\ 0 & 0 & \varepsilon_z \end{bmatrix},$$

where, as in $MoOCl_2$, the three principal components differ from each other (the wavelength dependence is omitted for brevity).

Following Ref. [13], by choosing the plane waves as the basis to write the electromagnetic fields:

$$\boldsymbol{E} = \boldsymbol{e}E_0 e^{\iota(\kappa_0 \boldsymbol{q}\cdot\boldsymbol{r} - \omega t)}, \qquad \text{(S1)}$$
$$\boldsymbol{H} = \boldsymbol{h}H_0 e^{\iota(\kappa_0 \boldsymbol{q}\cdot\boldsymbol{r} - \omega t)},$$

where $\boldsymbol{q} = (q_x, q_y, q_z)$ is the normalized wavevector (with respect to the free-space wavevector $\kappa_0 = 2\pi/\lambda_0$) and $\boldsymbol{e}$ ($\boldsymbol{h}$) represents the electric (magnetic) field unit vector, and substituting:

$$\boldsymbol{\nabla} x \boldsymbol{H} = \frac{1}{c}\frac{\partial \boldsymbol{D}}{\partial t} \text{ (with } \boldsymbol{D} = \varepsilon \boldsymbol{E}\text{)}$$

into the curl of:

$$\boldsymbol{\nabla} x \boldsymbol{E} = -\frac{1}{c}\frac{\partial \boldsymbol{H}}{\partial t},$$

we obtain a linear system of three equations for the components of $\boldsymbol{e}$, which has solutions only when the determinant is zero.

That condition provides the Fresnel's equation for an infinite biaxial crystal[13,14]:

$$-q_z^2\left[-q_z^2\varepsilon_z + \varepsilon_z(\varepsilon_x + \varepsilon_y) - q_x^2(\varepsilon_x + \varepsilon_z) - q_y^2(\varepsilon_y + \varepsilon_z)\right] + \left(\varepsilon_z - q_x^2 - q_y^2\right)\left(\varepsilon_x\varepsilon_y - q_x^2\varepsilon_x - q_y^2\varepsilon_y\right) = 0, \qquad \text{(S2)}$$

with solutions:

$$q_{o,e_z}^2 = -\frac{1}{2}\left\{\left[\frac{\varepsilon_x + \varepsilon_z}{\varepsilon_z}q_x^2 + \frac{\varepsilon_y + \varepsilon_z}{\varepsilon_z}q_y^2 - \left(\varepsilon_x + \varepsilon_y\right)\right] \pm \sqrt{D}\right\} \qquad \text{(S3)}$$
$$D = \left(\varepsilon_x - \varepsilon_y + \frac{\varepsilon_z - \varepsilon_x}{\varepsilon_z}q_x^2 - \frac{\varepsilon_z - \varepsilon_y}{\varepsilon_z}q_y^2\right)^2 + \frac{4(\varepsilon_z - \varepsilon_x)(\varepsilon_z - \varepsilon_y)}{\varepsilon_z^2}q_x^2 q_y^2.$$

Each solution represents the possible $q$-vectors in the bulk material at a given wavelength, forming a 3D surface in reciprocal space, i.e., the 3D isofrequency contours (IFCs), whose shape changes with wavelength according to the material dispersion. Since the solution space for the components of $\boldsymbol{e}$ is two-dimensional, there are two IFCs for each of the two possible polarization states, called ordinary and extraordinary modes[13]:

$$\boldsymbol{e_o} = \frac{1}{q}\begin{pmatrix} -q_y(1 - \Delta_1\Delta_z) \\ q_x \\ \mp q_x q_y q_{o_z}\Delta_1 \end{pmatrix}, \quad \boldsymbol{e_e} = \frac{1}{q}\begin{pmatrix} q_x\frac{\Delta_2 - q_y^2}{\Delta_x^e} \\ q_y \\ \frac{\Delta_2}{\mp q_{e_z}} \end{pmatrix}, \qquad \text{(S4)}$$

where $q = q_x^2 + q_y^2$ and

$$\Delta_1 = \frac{\Delta_x^o - q_x^2}{\Delta_z\Delta_x^o - q_x^2 q_{o_z}^2}, \qquad \Delta_2 = \frac{\Delta_x^e\Delta_y^e - q_x^2 q_y^2}{\Delta_x^e - q_x^2},$$
$$\Delta_x^{o,e} = \varepsilon_x - q_y^2 - q_{o,e_z}^2, \quad \Delta_y^{o,e} = \varepsilon_x - q_x^2 - q_{o,e_z}^2, \quad \Delta_z = \varepsilon_z - q_x^2 - q_y^2.$$

In case the biaxial crystal displays a highly dispersive response arising from strong material resonances interacting with light, these modes represent the polariton waves.

At a given wavelength, the properties of the IFCs for the ordinary and extraordinary modes change as a function of the in-plane wavevector depending on the value of the discriminant $D$ in Equation S3. In particular, there exist three possible types of waves: propagating, evanescent, and ghost[15]. The first two take place if $D > 0$, i.e., the solutions $q_{o,e_z}$ are real, and three cases are possible: $q_{o_z}$ and $q_{e_z}$ are both positive (2 propagating waves), both negative (2 evanescent waves) or one is positive and one negative (1 propagating and 1 evanescent wave). Finally, if $D < 0$, $q_{o_z}$ and $q_{e_z}$ are complex: in other words, they have both propagating (oscillatory) and evanescent (exponentially decaying) characters. Remarkably, this can happen also in absence of material losses, i.e., the dielectric permittivity is real. Such a condition defines the peculiar properties of the (bulk or volume-confined) ghost waves, and can be fulfilled only when[15]:

$$\min\{\varepsilon_x, \varepsilon_y\} < \varepsilon_z < \max\{\varepsilon_x, \varepsilon_y\}.$$

In $MoOCl_2$, this condition is satisfied over the whole frequency range explored in our work, since $\mathrm{Re}(\varepsilon_y) > \mathrm{Re}(\varepsilon_z)$. A necessary (but not sufficient) condition for bulk ghost waves to exist in a crystal is that all the diagonal elements of the permittivity tensor must be different (this is the case for biaxial $MoOCl_2$). This condition highlights the difference with the ghost modes recently observed in calcite. As calcite is a uniaxial crystal, the above condition cannot be satisfied, and ghost modes in the bulk crystal do not exist. However, ghost modes in calcite can appear as surface-confined waves at the interface between a semi-infinite off-axis cut uniaxial crystal and a dielectric and require $\varepsilon_\perp cos\theta^2 + \varepsilon_\parallel sin\theta^2 < 0$. Here $\varepsilon_\perp = (\varepsilon_x, \varepsilon_z)$ and $\varepsilon_\parallel = \varepsilon_y$ with $\theta$ being the angle defining the off axis cut.

For bulk $MoOCl_2$, we report in Fig. S4a the IFCs at $\lambda_0$ = 633 nm, together with their projection on the $q_{xy}$ plane (as in Fig. 1 of the main text). The occurrence of the previous cases is marked by the colormap, which can be better appreciated in the zoom of Fig. S4b, defining the existence domain of each solution type: blue for the two propagating waves, orange when only one propagating wave exist, light purple for two ghost waves (we only plot the real part of the wavevector, as we are interested in the propagation properties), and white if both modes are evanescent.

Accordingly, the boundaries of these domains on the $q_{xy}$ plane are set by the previous conditions on $q_{o,e_z}$. In the case of $MoOCl_2$, the IFCs corresponding to the 2D regions where only one propagating wave exists, having either isotropic or hyperbolic character (spheroidal and hyperboloidal purple surfaces, respectively), as described in the main text. These regions are delimited by the dotted and dashed black lines in Fig. S4b, defined by the equations[15]:

$$q_x^2 + q_y^2 = \varepsilon_z \text{ (isotropic)},$$

$$\frac{q_x^2}{\varepsilon_y} + \frac{q_y^2}{\varepsilon_x} = 1 \text{ (hyperbolic)}.$$

The former defines a circle of radius $\sqrt{\varepsilon_z}$ (= 1.9 at 633 nm), the latter a hyperbola with vertex at $q_x = \pm\sqrt{\varepsilon_y}$ (= 2.9 at 633 nm).

On the other hand, the portion of the $q_{xy}$ plane supporting ghost waves is defined by the lines[15]:

$$Aq_x \pm Bq_y \pm C = 0, \qquad \text{(S5)}$$

$$A = \sqrt{\frac{|\varepsilon_x - \varepsilon_z|}{\varepsilon_z}}, B = \sqrt{\frac{|\varepsilon_y - \varepsilon_z|}{\varepsilon_z}}, C = \sqrt{|\varepsilon_y - \varepsilon_x|},$$

which is also linked to open IFCs as for the hyperbolic waves, meaning that their wavevectors can be extremely large (and directional as well, as they are bound by parallel lines).

Finally, there is an area surrounded by the previous ones, which has a lenticular shape (blue region) and is characterized by the presence of two propagating waves. Such modes will be encountered also in the thin film limit.

The difference between the number and the type of allowed waves for each mode is represented in Fig. S4c, d, where we plot separately the ordinary and extraordinary waves, respectively, which are otherwise indistinguishable in Fig. S4a. We can notice that, in line with the colormap, isotropic and hyperbolic waves only exist in one case (extraordinary mode). Fig. S4c also helps to identify the IFCs of the lenticular waves, which are hidden by the ghost IFCs in Fig. S4a.

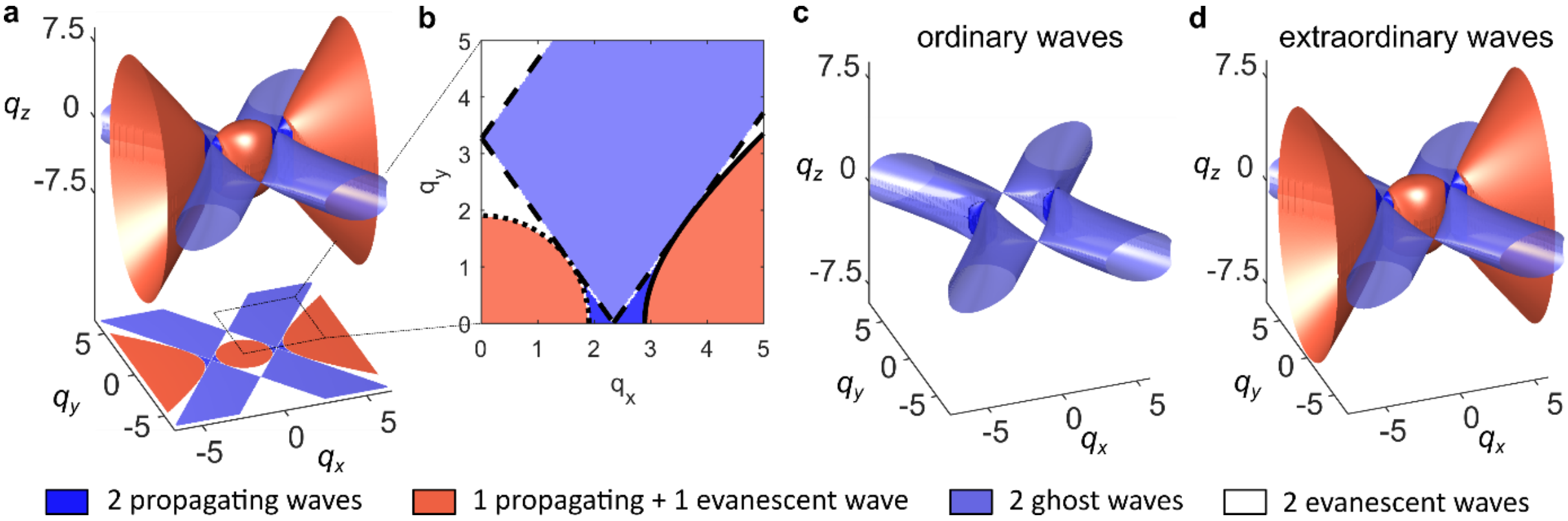

**Figure S4. Solutions of Fresnel's equation for $MoOCl_2$. a** IFCs (of both ordinary and extraordinary modes) at $\lambda_0$ = 633 nm with their in-plane projection. **b** Zoom of in-plane domains of the IFCs where spherical (orange), hyperbolic (also orange), lenticular (blue) and ghost modes (light purple) can be seen together with their boundaries (black lines). Only positive $q_x$ and $q_y$ are shown due to symmetry. **c** IFCs corresponding to the ordinary modes ($\boldsymbol{e_o}$), which have only lenticular or ghost character. **d** IFCs corresponding to the extraordinary modes ($\boldsymbol{e_e}$).

For a full description of volume-confined modes in bulk $MoOCl_2$, we complement Fig. S4a (or, equivalently, Fig. 1f of the main text) with analogous plots of the IFCs at different wavelengths across the hyperbolic window (see Fig. S2). Fig. S5a-d display the evolution of their shape at the four selected wavelengths: 830, 730, 633, 532 nm, respectively. As the wavelength decreases, the hyperbolic modes tend to spread less in any direction (as a consequence of the change in their asymptotes, discussed below), while the ghost modes modify the aperture of their IFCs in the $q_{xy}$ plane. In contrast, since $\varepsilon_z$ is almost constant in this range (see Fig. S2), the spheroidal IFC corresponding to in-plane isotropic modes is mostly unchanged.

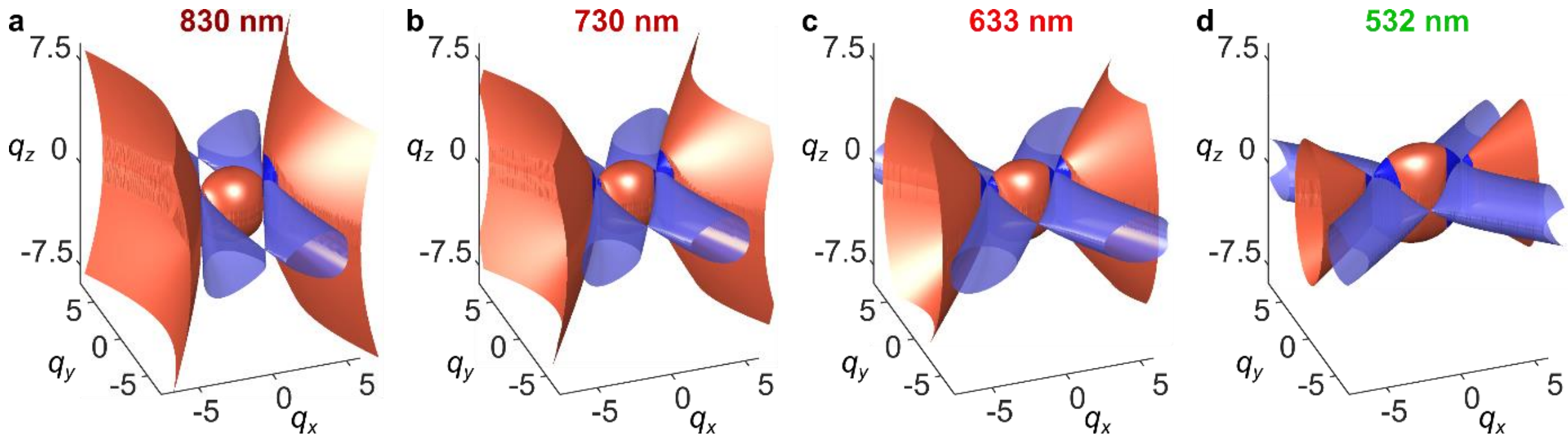

**Figure S5. Bulk $MoOCl_2$ isofrequency contours as a function of wavelength. a** $\lambda_0 = 830$ nm. **b** $\lambda_0 = 730$ nm. **c** $\lambda_0 = 633$ nm. **d** $\lambda_0 = 532$ nm. For more generality, only the extraordinary mode is plotted (the ordinary mode coincides with it for the ghost and lenticular waves).

In the previous analysis (and in Fig. 1f and 2a of the main text), we completely neglected the imaginary part of the dielectric tensor. As shown in Fig. S2, $MoOCl_2$ optical losses in the visible and near-infrared ranges are negligible, hence we do not expect any sharp variation in the IFCs calculated with complex $\varepsilon$. This is confirmed by Fig. S6, where we plot the real part of $q_{o_z}$ and $q_{e_z}$ at 633 nm. Here, the distinction between the modes based on the propagating or damped (i.e., evanescent) characters ceases to exist due to the complex nature of the wavevectors, hence the same colour is used for all the solutions. Interestingly, the imaginary part of $\varepsilon$ only induces a minor change with respect to the real case, i.e., the spheroidal, hyperbolic and ghost IFCs slightly merge their boundaries, although with no consequences on the individual shapes. This justifies the validity of our analytical approach, both for the bulk crystal and the thin film (shown below), where the material is treated as lossless.

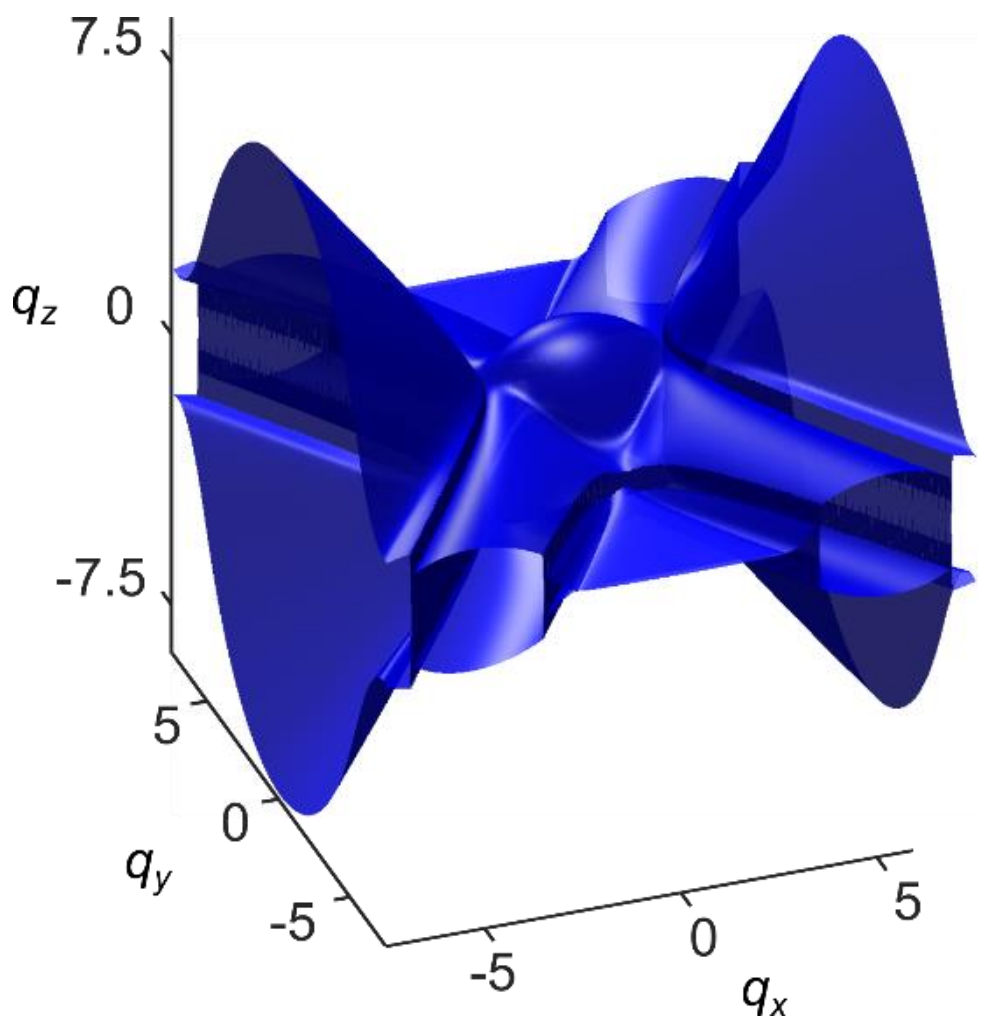

**Figure S6. Bulk $MoOCl_2$ isofrequency contours computed with complex permittivity ($\lambda_0$ = 633 nm).** For simplicity, only the extraordinary mode is plotted. Same colour is used for hyperbolic, ghost and spherical waves, as they all possess both propagating and evanescent characters.

One last observation concerns the asymptotes of the hyperbolic and ghost IFCs, which characterize the unidirectional propagation of these waves in the bulk crystal, both in- and out-of-plane. The asymptotic in-plane propagation of hyperbolic waves can be found from Equation S2, by setting $q_z = 0$ (resulting in the 2D IFCs of Fig. S7a) and allowing $q_{x,y} \to \infty$ (hence, focusing on large wavevectors as in the top-right red box in Fig. S7a, shown in Fig. S7b), which yield the following directions for the in-plane wavevector (with respect to the $x$-axis):

$$\pm\arctan\left(\frac{q_y}{q_x}\right) = \pm\arctan\left(\sqrt{-\frac{\varepsilon_x}{\varepsilon_y}}\right), \tag{S6}$$

hence, since the propagation direction is defined by the Poynting vector, whose direction in real space is given by the normal to the IFCs (Fig. S7b), the in-plane directional hyperbolic rays are defined by the angles:

$$\pm\alpha_H = \pm\left[\frac{\pi}{2} - \arctan\left(\sqrt{-\frac{\varepsilon_x}{\varepsilon_y}}\right)\right], \tag{S7}$$

equal to $\approx 50°$ at $\lambda_0$ = 633 nm. Such angle is relevant for the present discussion since it describes the asymptotic in-plane aperture ($2\alpha_H$) of hyperbolic polariton rays in bulk $MoOCl_2$, thus defining the maximum angle at which we can observe these waves in our simulations (Fig. 2b, c of the main text). Similarly, as bulk ghost modes are bound by parallel lines in the $q_{xy}$ plane, their in-plane asymptotes in real space (with respect to the $x$-axis, Fig. S7b) can be computed from the angular coefficient of those lines (Equation S5):

$$\pm\alpha_G = \pm\left[\frac{\pi}{2} - \arctan\left(\frac{A}{B}\right)\right] = \pm\left[\frac{\pi}{2} - \arctan\left(\sqrt{\frac{|\varepsilon_x-\varepsilon_z|}{|\varepsilon_y-\varepsilon_z|}}\right)\right],$$

equal to 36° for $\lambda_0$ = 633 nm. This is the angle at which we monitored the simulated fields of bulk plasmon polaritons in a thick $MoOCl_2$ slab (Fig. 2b of the main text), showing the presence of ghost waves that propagate almost planarly (see discussion of Fig. S9).

Notably, in real experiments the actual polariton propagation direction depends on the excited wavevectors, which is in turns related to the size of the object used to launch/probe these modes. In other words, if the momenta that the object can supply are large compared to the free space wavevector, i.e., $q \gg 1$, the in-plane direction is fixed by the asymptotes of the hyperbolic ($\alpha_H$) and ghost ($\alpha_G$) IFCs (Fig. S7b). Otherwise, if $q \sim 1$ (i.e., magenta box in Fig. S7a), the propagation of hyperbolic waves changes according to the curvature of their IFCs as sketched in Fig. S7c: in this case the ghost Poynting vector ($\boldsymbol{S_G}$) is unchanged, while the hyperbolic one ($\boldsymbol{S_H}$) forms a smaller angle with the $x$-axis. Hence, using antennas of various sizes can in principle affect the observed propagation (see also Supplementary Information S8).

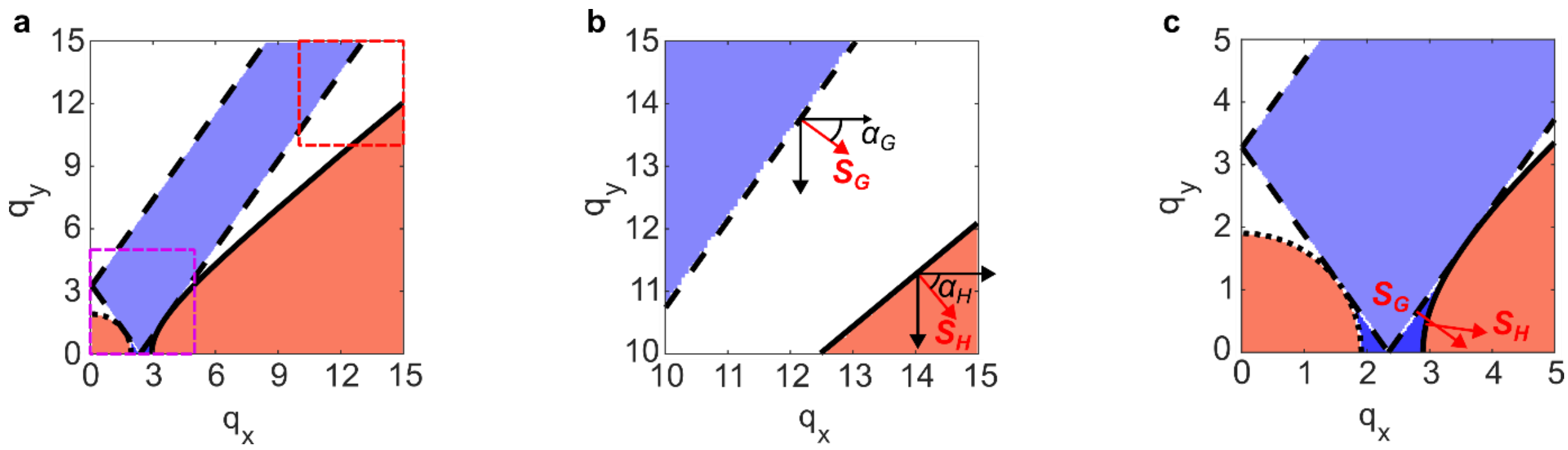

**Figure S7. In-plane asymptotes of bulk polaritons in $MoOCl_2$ ($\lambda_0$ = 633 nm). a** In-plane projection of bulk IFCs with boundaries separating the different solution's domains. Dashed squares indicate regions of small (magenta) and large (red) wavevectors. **b** Zoom of dashed red square region in **a** to visualize asymptotic in-plane behaviour of ghost and hyperbolic waves. Poynting vectors ($\boldsymbol{S_G}, \boldsymbol{S_H}$) and the propagation directions ($\alpha_G$, $\alpha_H$) are also sketched. **c** Zoom of dashed magenta region in **a** to highlight ghost and hyperbolic waves direction at low wavevectors.

When it comes to the out-of-plane propagation, it is useful to discuss hyperbolic and ghost waves separately, as the $z$-confinement is due to different mechanisms. As shown in Fig. S8a, the hyperbolic IFCs are characterized also by an out-of-plane asymptote, whose direction changes as different parts of the IFCs are probed, setting the specific out-of-plane ray-like direction of hyperbolic polaritons in real-space. This is determined by the crystal plane at which light propagation is observed, as defined by the angle $\theta$ of Fig. 2b of the main text. For a given $\theta$, the associated part of the IFCs (i.e., the hyperbolic waves involved) is determined by the ensemble of points whose Poynting vectors satisfy $\theta = \arctan(S_y/S_x)$, fixing the in-plane propagation direction, while the out-of-plane direction is defined by $\arctan(S_z/S_x)$. Since the hyperbolic IFCs are linear in the long-wavevector limit, this locus of points asymptotically corresponds to the intersection of the IFCs with a $q_x$-$q_z$ plane rotated by an angle $\beta$ around the $q_z$-axis (blue line in Fig. S8a, where the red arrow sketches a representative Poynting vector in the asymptotic part of the IFCs). Note that the angles $\theta$ (in real space) and $\beta$ (in momentum space) are generally not the same. Nevertheless, by looking at Fig. S8b, one can see that an increase in $\beta$ corresponds to an increase in $S_y/S_x$, which in turn indicates an increase in $\theta$. For example, at $\beta = 0$ we have $S_y = 0$ (top arrow in Fig. S8b), implying $\theta = 0$, which increases as the Poynting vector approaches the $q_x$-$q_y$ plane. Therefore, bearing in mind this relationship between $\theta$ and $\beta$, Fig. S8b tells us that an increase in $\beta$ corresponds to a decrease of the ratio $S_z/S_x$, hence a decreasing out-of-plane opening angle of the hyperbolic rays in real-space. This is in line with the observed out-of-plane behaviour of the hyperbolic wavefronts in Fig. 2b of the main text.

We extracted the dependence of the out-of-plane direction on $\beta$ by computing the normal to the IFCs at a point $(q_x, q_y)$ along different directions $\beta$, as plotted in Fig. S8c. Since hyperbolic IFCs have $q_y$-$q_z$ sections that are not circular, as the aperture of the hyperboloid changes in this plane ($\varepsilon_x \neq \varepsilon_y \neq \varepsilon_z$), the shape of the plot reflects this asymmetry, with a steeper decrease compared to a circle (see dashed line) as $\beta$ is increased. Note that the out-of-plane angle drops at zero (in-plane propagation) before the predicted $\beta = 40°$ (Equation S6), because of the finite values of $(q_x, q_y)$ at which the Poynting vector is calculated.

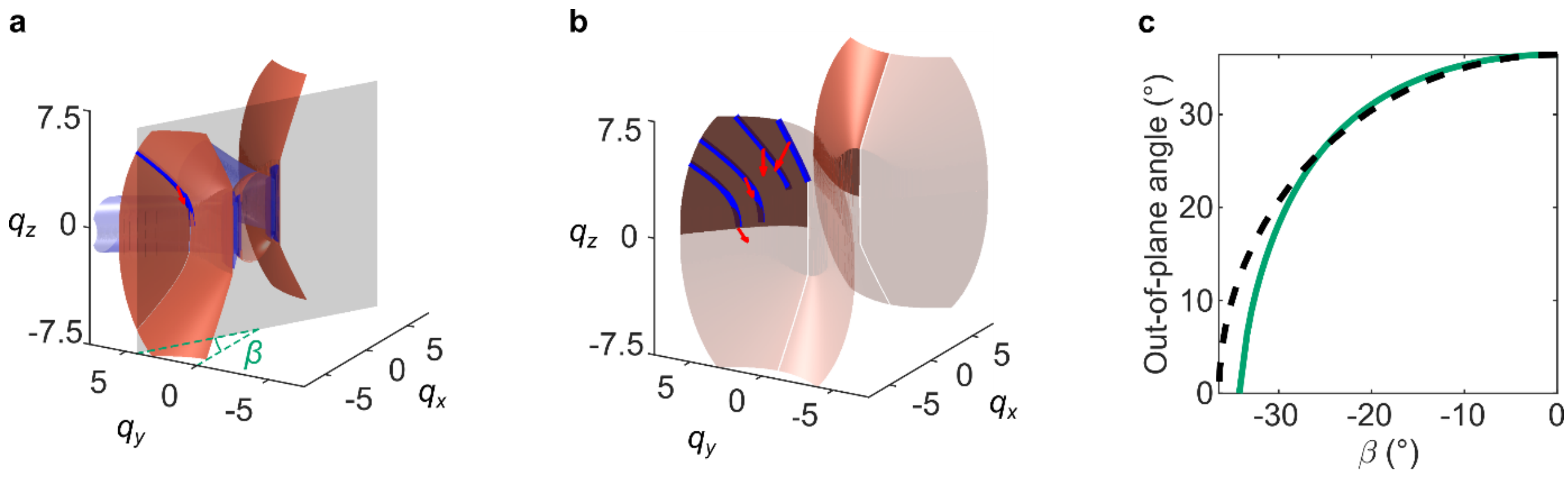

**Figure S8. Out-of-plane asymptotes of bulk hyperbolic polaritons in $MoOCl_2$ ($\lambda_0$ = 633 nm).** **a** Section of bulk IFCs for $q_y > 0$ intersected by plane (grey) at angle $\beta$ representing a specific in-plane direction. Red arrow (orthogonal to IFCs) represents the Poynting vector associated with asymptotic region of the IFCs. **b** Zoom of dashed red square region in **a** to visualize asymptotic in-plane behaviour of ghost and hyperbolic waves. Poynting vectors ($\boldsymbol{S_G}$, $\boldsymbol{S_H}$) and the propagation directions ($\alpha_G$, $\alpha_H$) are also sketched. **c** Zoom of dashed magenta region in **a** to highlight ghost and hyperbolic waves direction at low wavevectors.

Regarding ghost polaritons, while their in plane (x-y) directionality comes from the straight IFCs, their out-of-plane confinement (as seen in Fig. 2b of the main text) is due to an interplay between the shape of ghost IFCs and the losses that affect different parts of them. As for hyperbolic waves, to understand the z-confinement we intersect ghost IFCs with a $q_x$-$q_z$ plane rotated by an angle $\beta$ around the $q_z$-axis, as in Fig. S9a. Here, the choice of $\beta$ is limited by the direction of the parallel lines that confine ghost modes in the plane (Equation S5). An example of a section of ghost IFCs is sketched in Fig. S9b (yellow lines), in the case the angle is close to these lines' angular coefficient ($= \pm\arctan(A/B)$). We can see that the part at small wavevectors (dashed black lines) is almost vertical, while the asymptotic behaviour is mainly horizontal. Since the propagation of ghost waves in real space follows the shape of the Fourier transform of the IFCs, which is sketched in Fig. S9c, we find that two principal propagation directions are possible: horizontal (along the in-plane asymptote $\alpha_G$) and vertical. This can also be seen by the directions of the Poynting vectors (red arrows) in Fig. S9b.
However, it turns out that ghost waves propagating in the $z$-direction are strongly damped compared to in-plane waves (dashed black lines). Indeed, if we consider the portions of ghost IFCs characterized by a $q_z$ whose real part ($\Re(q_z)$ in Fig. S9d) is larger compared to the imaginary part ($\Im(q_z)$ in Fig. S9e), we clearly see that this happens only for the regions at small wavevectors, as witnessed by the colormap of Fig. S9f, where $\Re(q_z) - \Im(q_z)$ is plotted. These regions correspond to horizontal waves (see Fig. S9b, c), in line with the observation of Fig. 2b of the main text.

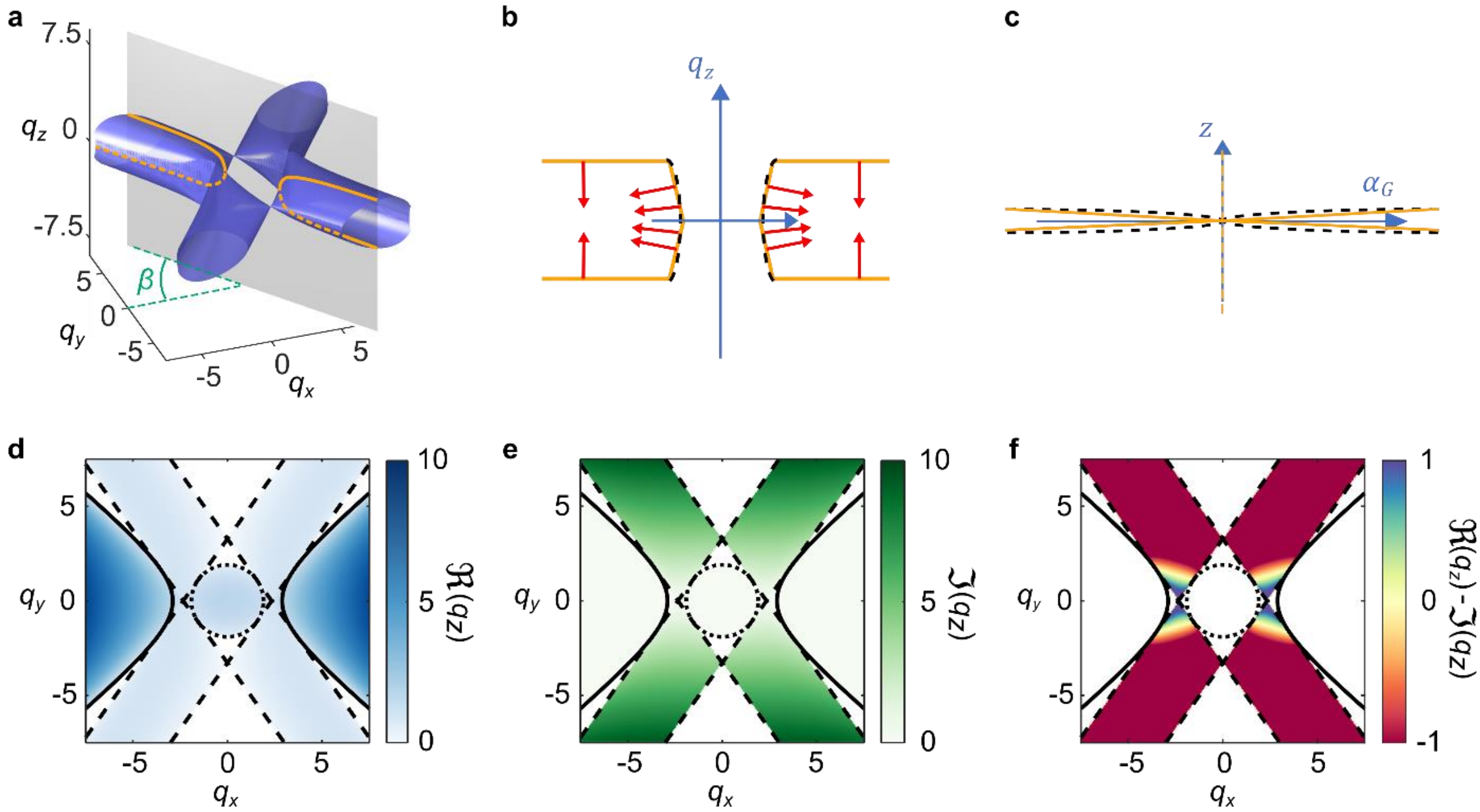

**Figure S9. Out-of-plane asymptotes of bulk ghost polaritons in $MoOCl_2$ ($\lambda_0$ = 633 nm).** **a** Ghost IFCs intersected by plane (grey) representing a specific in-plane $\beta$ direction. **b** Representation of the out-of-plane section of ghost IFCs (yellow) as given by the intersection in **a**. Dashed line represents their shape at small wavevectors. Red arrows sketch Poynting vectors at different points. **c** Representation of Fourier transform of **b**. Vertical line at the origin represents the Fourier components of horizontal asymptotes in **b**, while dashed line relates to the contribution from dashed line in **b**. **d**, **e** In-plane projection of the real and imaginary parts of the IFCs, respectively, as in Fig. 2a of the main text. **f** Difference between **d** and **e**, i.e., $\Re(q_z) - \Im(q_z)$.

## Thin films of $MoOCl_2$

Many practical experiments, such as the s-SNOM imaging discussed in this manuscript, involve a crystal of finite thickness ($d$) sandwiched by two dielectric media (both much thicker than $d$, hence approximated as semi-infinite) with permittivities $\epsilon_1$ and $\epsilon_3$, for the upper and lower half-spaces, respectively. The solution of Maxwell's equations for a biaxial crystal in this case is rather complex, involving an 8x8 linear system,[16] which is a formidable task even from a computational point of view.[15] Nevertheless, an analytical solution can still be derived in few limit cases. Of particular interest in the framework of thin van der Waals crystals, such as hexagonal boron nitride, $MoO_3$ and $MoOCl_2$, is the vanishing thickness limit, where the thickness is considered much smaller than the free-space wavelength: $\kappa_0 d = 2\pi\, d/\lambda_0 \ll 1$. Skipping the algebra, for which we refer to Ref. [13], the dispersion relation for electromagnetic modes in this approximation is found by:

$$\left\{\frac{\kappa_0 d\epsilon_x}{2\iota}q_y^2+\frac{\kappa_0 d\epsilon_y}{2\iota}q_x^2+\frac{q_x^2+q_y^2}{2}\left(\iota\sqrt{q_x^2+q_y^2-\epsilon_1}+\iota\sqrt{q_x^2+q_y^2-\epsilon_3}\right)\right\}$$
$$*\left\{\frac{\kappa_0 d\epsilon_x}{2\iota}q_x^2+\frac{\kappa_0 d\epsilon_y}{2\iota}q_y^2+\frac{q_x^2+q_y^2}{2}\left(\frac{\epsilon_1}{\iota\sqrt{q_x^2+q_y^2-\epsilon_1}}+\frac{\epsilon_3}{\iota\sqrt{q_x^2+q_y^2-\epsilon_3}}\right)\right\}$$
$$=$$

$$= -q_x^2 q_y^2\frac{\kappa_0^2 d^2}{4}\left(\epsilon_x-\epsilon_y\right)^2 \qquad \text{(S8)}$$

that provides the values of the in-plane wavevector $(q_x, q_y)$ as a function of wavelength. From this equation the out-of-plane wavevector can be also retrieved through Equation S3 (still valid within the biaxial crystal volume). Similar to the bulk crystal, also here we can study the asymptotes of the IFCs from Equation S8 upon setting $q_{x,y} \to \infty$ and retaining only the higher order terms. As expected, the asymptotes for the thin film coincide with the bulk ones extracted from the Fresnel equation (Equation S2) for the hyperbolic modes.[13] This is why we used Equation S8 to compute the aperture angle of hyperbolic rays in Fig. 4m of the main text.

While, for instance, in the case of mid-infrared (phonon-)polaritons in hBN or $MoO_3$, the condition $\kappa_0 d \ll 1$ is fulfilled for most thicknesses of as-exfoliated flakes, for visible plasmon-polaritons in $MoOCl_2$ this is not always true. In particular, already around a thickness of 80-90 nm, $\kappa_0 d$ reaches unity at $\lambda_0$ = 532 nm. To check for the validity of the analytical dispersion relation as a function of both the thickness and the free-space wavelength (across the explored hyperbolic window of $MoOCl_2$: $\lambda_0$ = 830-532 nm), we compare in Fig. S10 the solutions of Equation S8 (solid lines) and the transfer matrix calculations (superimposed colormap). In our case, $\epsilon_1 = 1$ (air), and $\epsilon_3 = \epsilon_{SiO_2}$. A good agreement is obtained for thicknesses of 20 and 40 nm (this is the thickness of the thin film characterized with s-SNOM in the main text, see Fig. 4), where the latter shows only minor deviations at shorter wavelengths. In contrast, large deviations between the analytical IFCs and the numerical calculations are observed for larger thicknesses (> 100 nm). At the same time, for a fixed thickness, the agreement between theory and simulations degrades as the wavelength is decreased (at longer wavelengths $\kappa 0$ is reduced). To visualize the evolution of the IFCs with $\lambda_0$, we report in Figure S10o the analitically calculated IFCs for a 40 nm film. Here, we can appreciate the variation of the hyperbolic aperture angle $(\pi/2-\alpha_H)$, which increases with the wavelength in momentum space. In particular, the hyperbolic rays' aperture angle in real space $(\alpha_H)$ decreases with $\lambda_0$, according to the previous discussion of Equation S8.

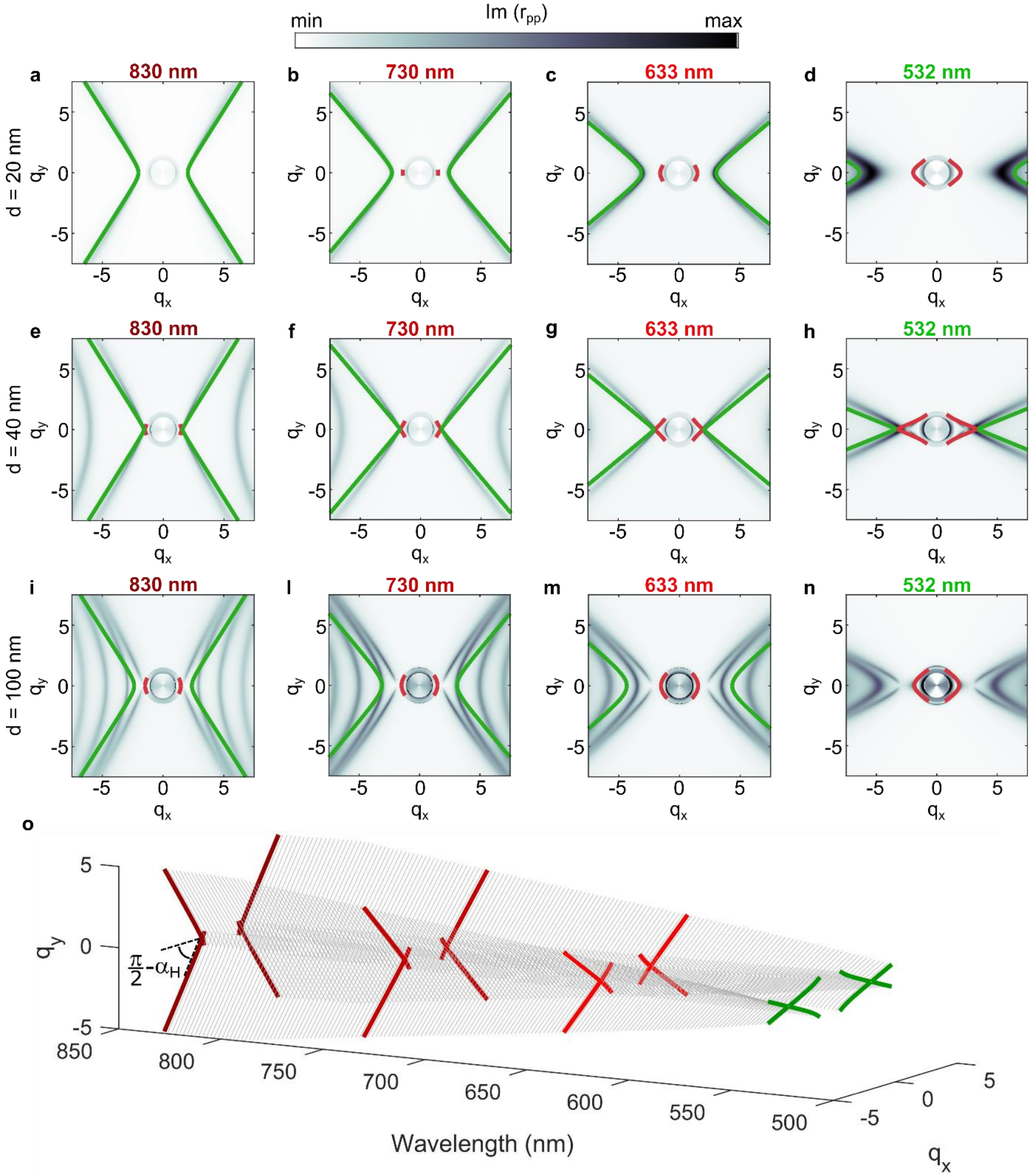

**Figure S10. Comparison of IFCs (of extraordinary modes only) calculated from analytical equation (solid lines) and transfer matrix method (colormap) for thin $MoOCl_2$ (thickness *d*) as a function of wavelength ($\lambda_0$ = 830, 730, 633, 532 nm). a-d** *d*=20 nm. **e-h** *d*=40 nm. **i-n** *d*=100 nm. Green lines: hyperbolic modes. Red lines: lenticular modes. **o** Evolution of the in-plane IFCs (d = 40 nm) when varying the illumination wavelength. The experimental wavelengths (Fig. 4 of main text) are highlighted with the corresponding colour. The IFCs half aperture angle is complementary to $\alpha_H$.

The consistency of the analytical formula with simulations for $d$ = 40 nm allows for a further step to interpret the experimental results of Fig. 4d-g of the main text. Specifically, there we noticed the presence of interference-like effects in the measured optical s-SNOM signal, which are particularly evident in the image at $\lambda_0$ = 633 nm (Fig. 4c). Regarding the origin of these features, we ascribe them to the concurrent propagation of hyperbolic and lenticular modes. To support this hypothesis, we analyse individually the in-plane ($x$-$y$) electric field ($\Re(E_y)$) profiles of both hyperbolic (Fig. S11a) and lenticular (Fig. S11b) modes at $\lambda_0$ = 633 nm, computed thanks to the knowledge of their analytical relations between the in-plane components of the wavevector (whose values are plug into Equations S1 and S4, in the case of extraordinary mode solution, which is the only possible one for hyperbolic waves). From Fig. S11a, b we can appreciate that both waves display a directional propagation along different in-plane directions, where each wavefront is tilted at a different angle with respect to each propagation direction. As a result, in case both modes exist at the same time (such as at $\lambda_0$ = 633 nm), their superposition is characterized by interference fringes forming a cross-shape centred around the common source (Fig. S11c), in line with the observed real-space propagation of PPs in Fig. 4f. Moreover, the 2D Fourier transform of Fig. S11c, reported in Fig. S11d, is in good agreement with the experimental data of Fig. 4j.

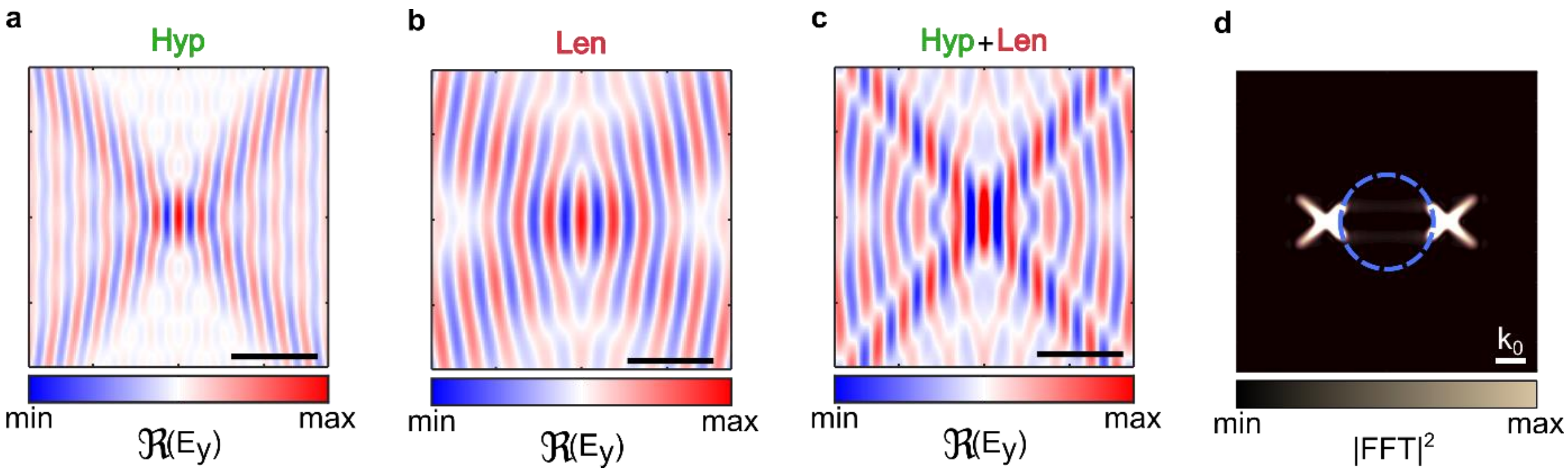

**Figure S11. Origin of interference-like features in s-SNOM imaging of polaritons in thin film $MoOCl_2$ ($\lambda_0$ = 633 nm, $d$ = 40 nm).** **a** In-plane distribution ($x$-$y$) of electric field $E_y$ of hyperbolic modes (resulting from all wavevectors of green lines in Fig. S10g). **b** Same for lenticular modes (wavevectors corresponding to red lines in Fig. S10g). **c** Sum of hyperbolic and lenticular electric fields of panels **a** and **b**. **d** 2D (square modulus of) Fourier transform of panel **c**. Dashed line represents $SiO_2$ light-cone. Black scale bars: 1 $\mu$m.

**Comparison of the modes supported in $MoOCl_2$ and calcite**

Here we summarize the main differences and similarities in the modes supported by $MoOCl_2$ and calcite.

In the investigated frequency range, $MoOCl_2$ is a biaxial crystal with $\varepsilon_x < 0$ and $\varepsilon_{y,z} > 0$. The bulk crystal (i.e., infinite slab) supports ghost modes, featuring a cross-shaped in-plane dispersion, that can propagate inside the volume of an infinite slab (see Fig. 2b of the main text). The condition for their existence is simply:

$$\min\{\varepsilon_x, \varepsilon_y\} < \varepsilon_z < \max\{\varepsilon_x, \varepsilon_y\}.$$

Calcite, instead, is a uniaxial crystal with two equal components of the dielectric function. In the higher Reststrahlen (RS) band, $\varepsilon_x = \varepsilon_z < 0$ and $\varepsilon_y > 0$. The previous condition for bulk ghost modes cannot be satisfied in a uniaxial crystal. Nevertheless, ghost modes can still exist in calcite as surface waves (also called Dyakonov waves),[17] propagating along the interface between the semi-infinite uniaxial slab and an isotropic dielectric. They are characterized by an in-plane hyperbolic dispersion and the condition for their existence is:[10]

$$\varepsilon_{x,z} cos\theta^2 + \varepsilon_y sin\theta^2 < 0,$$

where $\theta$ indicates the off axis cut of the interface with respect to the crystal axis. In the lower RS band, where the signs of the tensor elements are swapped ($\varepsilon_x = \varepsilon_z > 0$ and $\varepsilon_y < 0$), a semi-infinite slab of calcite instead supports leaky modes at the interface with an isotropic dielectric. These modes, characterized by closed IFCs,[18] are named leaky as their dispersion crosses the material light cone in the z-direction (defined by $n_z$) Hence, they are evanescent in the upper plane (i.e., in the dielectric), but can propagate inside calcite. These modes exist independently of the interface cut being off axis. Moreover, the requirements on the sign of the tensor elements imply that ghost and leaky modes appear in distinct RS bands at separate frequencies.
Given the presence of surface modes at the interface between calcite and an isotropic dielectric, one could wonder if similar surface modes exist in biaxial $MoOCl_2$. These are known as Dyakonov surface waves and the condition for their existence is:[19]

$$\min\{\varepsilon_x, \varepsilon_y\} \leq \varepsilon_z < \varepsilon_0 < \max\{\varepsilon_x, \varepsilon_y\}$$

where $\varepsilon_0$ is the dielectric function of the homogeneous medium. In the upper RS band of calcite, this condition is fulfilled as $\varepsilon_x = \varepsilon_z < 0$ and $\varepsilon_y > \varepsilon_0 = 1$. Ghost modes coincide with Dyakonov modes for the specific calcite cut at $\theta = 0°$, where the out-of-plane wavevector becomes purely imaginary (for $\theta > 0°$, the out-of-plane wavevector becomes complex). In $MoOCl_2$, instead, the Dyakonov condition is never fulfilled in the investigated frequency range since $\varepsilon_z > \varepsilon_0 = 1$ (air), and thus we do not expect to have surface modes at the interface. Notably, the surface ghost waves described in Ref. [[19]] are distinct from the ones of calcite, as they do not require a slanted cut of the crystal. Instead, they acquire ghost-feature when their in-plane momenta $q_x, q_y$ lie inside the **bulk**-ghost region defined by equations S3 and S5.

In thin films of $MoOCl_2$, we also observe modes that have an in-plane lenticular dispersion like the leaky modes observed in calcite in the lower RS band. As in calcite, these modes exist inside the material light cone, but their momentum is also lower bound by the refractive index of the substrate. They can exist even if Dyakonov waves are forbidden, as they are not purely interfacial modes (i.e., purely imaginary wavevector on both sides of the interface). Indeed, they exist within $MoOCl_2$ light cone with purely real wavevector (inside the anisotropic material). Since the conditions for leaky modes (i.e., $\varepsilon_y, \varepsilon_z > 0$ and $\varepsilon_x < 0$) re satisfied for $MoOCl_2$, we identify these modes as the thin-film version of the leaky-modes observed in semi-infinite slabs of calcite. Since the two conditions for bulk ghost modes and leaky modes are independent, ghost waves exist at the same frequency of lenticular modes in $MoOCl_2$, differently to what happens in calcite.

## S3 – Far-field optical characterization of the flakes

### Optical images from different polarization conditions

The anisotropic response of the exfoliated flakes is readily visible in the optical images taken with polarized light, as shown in Fig. 4 of the main text and in Fig. S12 below. The images were taken in an optical microscope by orienting the polarization of the illuminating light along the optical axes of the crystal, as depicted in the corresponding schematics in Fig. S12a, b.
Interestingly, we measured a non-zero reflection also in the cross-polarization condition (see Fig. S12c). This measurement was taken by rotating the polarization of the impinging light off-axis by 45° and by adding another polarizer in the collection arm rotated by 90° with respect to the illumination. Note that, if the polarization of the illuminating light is rotated along one of the crystal axes (as in Fig. S12a or b), no signal is visible in cross-polarization. The presence of strong cross-polarized reflection which disappears if the polarization is parallel to the crystal axes is signature of the birefringence of $MoOCl_2$, which is another aspect linked to its anisotropy.
In uniaxial media, birefringence causes light to travel with a phase retardation between the two polarizations parallel to the optical axes. This effect causes an effective rotation of the polarization of light when passing through the birefringent material. In general, birefringence can be quantified as the difference of refractive indexes in the plane perpendicular to the propagation direction of light:

$$|\Delta n|_{\parallel,\perp} = \left|\Re(n_x) - \Re\left(n_{y,z}\right)\right|$$

Where ∥ (⊥) refers to the in-plane $x - y$ (out-of-plane $x - z$) birefringence. We computed the in-plane and out-of-plane birefringence values by employing the dielectric functions of $MoOCl_2$ (see Section S1). The results (Fig. S12d) show that $MoOCl_2$ has a natural birefringence comparable and even larger than the anisotropic van der Waals materials studied before[20].

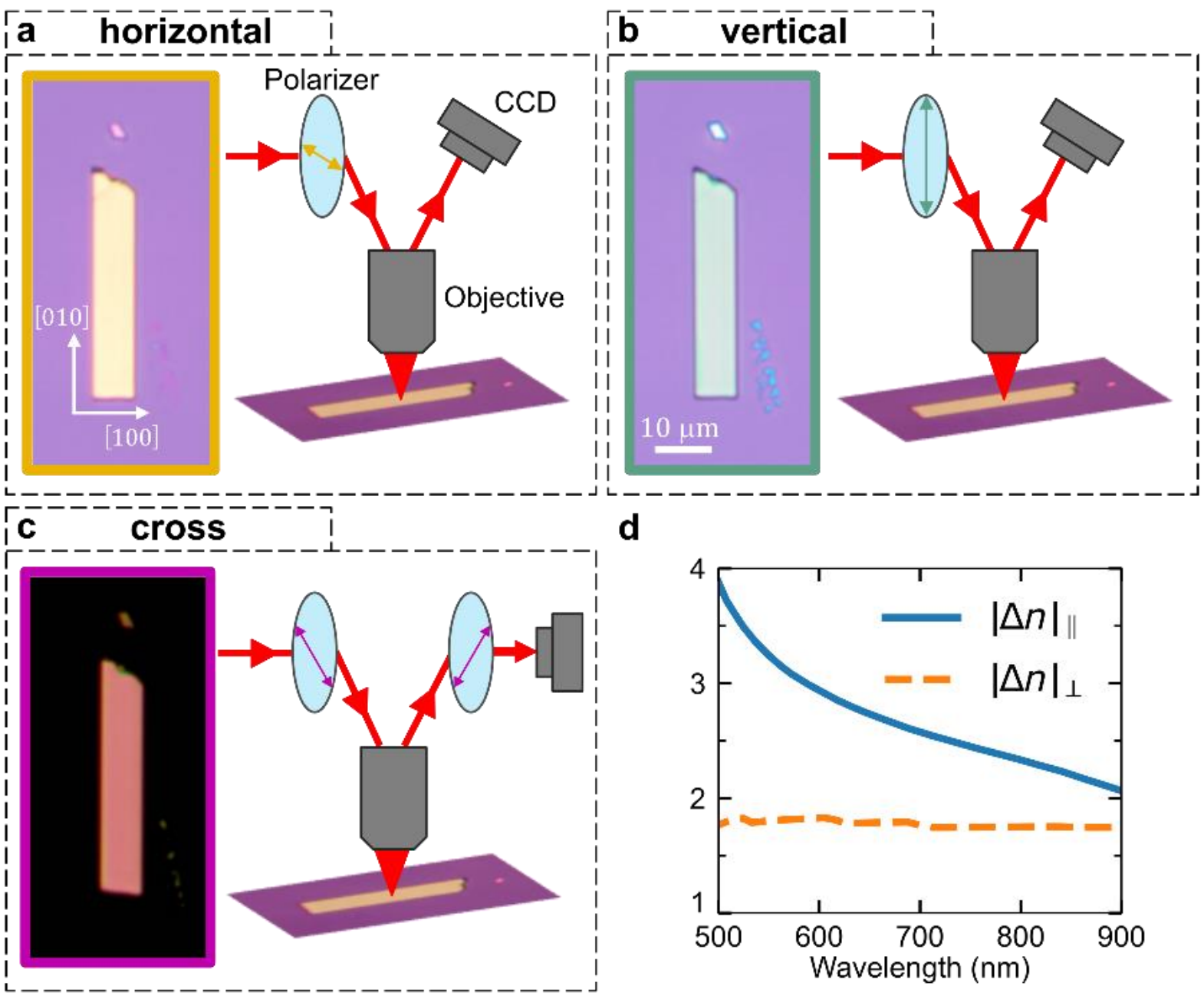

**Figure S12 – Optical images of an MoOCl2 flake upon polarized illumination. a-c** optical images (left) and sketch of the measurement setup (right) for three illumination condition: **a** polarization along the $|x\rangle$ axis, **b** polarization along the $|y\rangle$ axis and **c** off-axis cross polarization. **d** computed in-plane (blue solid line) and out-of-plane (orange dashed line) birefringence for $MoOCl_2$ in the frequency region of interest.

**Far-field reflectance measurements**

Fig. S13a shows optical images of different $MoOCl_2$ flakes exfoliated on a $SiO_2$ substrate. Remarkably, many of the flakes breaks in a regular rectangular shape, which we demonstrated to be oriented as the crystal axes (short side along [100] and long side along [010], see also main text and AFM analysis below).

We collected different reflectance spectra for flakes of various thickness with light polarized along the $|x\rangle$ (Fig. S13b) and the $|y\rangle$ (Fig. S13c) directions. The reflectivity of F2 was used in the main text (Fig. 4). The reflectivity of light polarized along the metallic $|x\rangle$ axis show very little variation with the thickness, as most of the light is reflected at the surface of the crystal. The reflectance along the $|y\rangle$ axis shows instead oscillations associated with multiple reflection between the crystal interfaces as inside a dielectric Fabry-Perot (FP) cavity with refractive index $n_y$. The periodicity of the oscillations increases when reducing the thickness, as expected. This is another confirmation of the anisotropic nature of $MoOCl_2$.

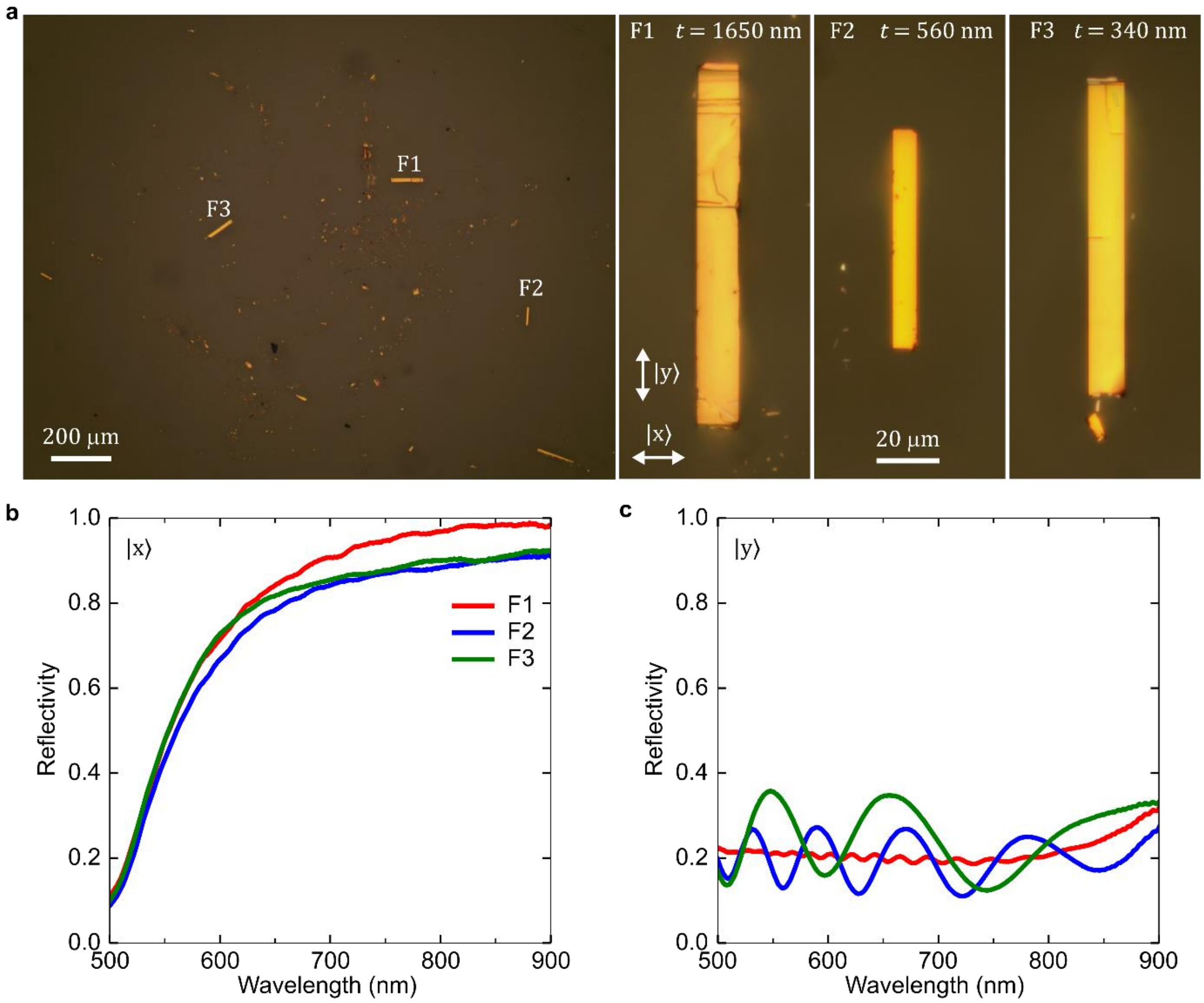

**Figure S13. Far-field optical characterization of different $MoOCl_2$ flakes. a** Optical images of different $MoOCl_2$ flakes exfoliated on $SiO_2$ and illuminated with unpolarized light. **b** Reflectivity spectra obtained with a polarized source along the $|x\rangle$ direction and **c** along the $|y\rangle$ direction on exfoliated flakes of different thickness.

## S4 – Atomic resolution AFM

The atomic resolution AFM image shown in Fig. 4 of the main text was post-processed as reported in Methods to remove the undesired noise and drift accumulated during the measurements. The atomic resolution AFM image was taken with the flake oriented as in Fig. S14a, so that the [100] and [010] axes of the crystal coincide with the fast scan and slow scan of the measurement, respectively. The image shown in Fig. S14b is a 4 x 6 $nm^2$ raw AFM data levelled by mean plane subtraction in Gwyddion (version 2.65). A clear atomic periodicity is visible, emerging from the repulsive interaction (Pauli repulsion) between the tip and the sample atoms[21,22]. By taking line profiles of the data, two different periodicities are seen for vertical and horizontal directions in the crystal. In particular, vertical line profiles taken in between the maxima of the map (green lines 1 and 2 in Fig. S14b, c) show a period with two different intensity maxima, which we tentatively associated with the Mo and Cl atoms. Vertical profiles on the maxima (green lines 3 and 4) provide a less pronounced modulation of the intensity. On the other hand, horizontal profiles show just a single intensity maxima value in all the sites both for profiles on the maxima (tentatively associated to Mo atoms, see lines 5 and 6) and in between the maxima (lines 7 and 8).

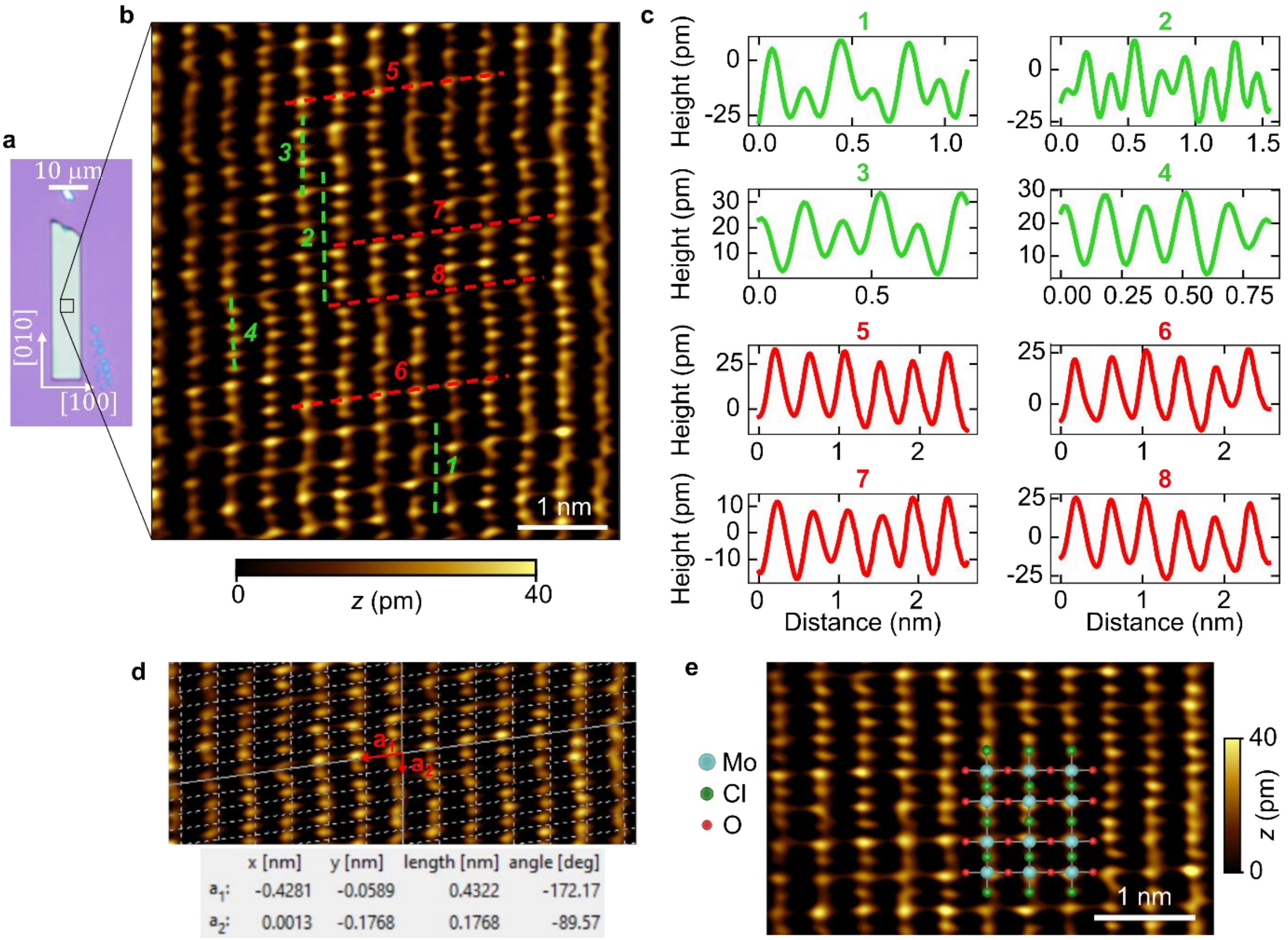

**Figure S14. Analysis of the AFM measurements. a** Optical image of the $MoOCl_2$ flake measured at the AFM. **b** Raw atomic resolution AFM image. Eight line profiles are shown. **c** Line profiles of the AFM measurement shown in **b. d** Analysis of the AFM image shown in **b**, obtained with the Lattice tool of the software Gwyddion. The red arrows indicate the extracted lattice vectors. **e** AFM image after drift correction. The expected $MoOCl_2$ crystal structure is superimposed for comparison.

We extracted the lattice base vectors of the rectangular lattice shown in the measurement with the Lattice tool of Gwyddion from the unprocessed data, as shown in Fig. S14d, obtaining the values $a_1 = 4.32$ Å and $a_2 = 1.77$ Å. From these values, we can compute the in-plane unit vectors $a$ and $b$ of the crystal lattice (see Fig. 1 in the main text) as:

$$a = a_1 \qquad b = 4a_2$$

The computed values of $a = 4.32$ Å and $b = 7.08$ Å are in good agreement with the values reported in Ref. [23] for bulk $MoOCl_2$ ($a = 3.8$ Å, $b = 6.5$ Å). Thanks to these considerations, we conclude that the AFM is probing the surface crystal structure of $MoOCl_2$. In fact, a good agreement is found between the theoretical lattice structure and the AFM measurement after the removal (by Gwyddion) of the drift distortion accumulated during the measurement, as shown in Fig. S14e.

## S5 – Transfer Matrix calculations

### Transition from bulk to thin film modes

Since the hyperbolic excitations in the thin film are formed by bulk PPs reflected at the material interfaces, as explained in the main text, it is interesting to follow the transition from bulk to thin film modes. To this end, we calculate the modes dispersion through the transfer matrix method[24] along the $x$ [100] and $y$ [010] directions for various thicknesses at $\lambda_0 = 633$ nm (Fig. S15). For each direction we mark with dashed lines the values of the refractive indexes of the substrate ($n_{sub} = 1.5$, blue line) and of the out-of-plane component of $MoOCl_2$ ($n_z = 1.9$, purple line). In the [010] direction a series of peaks for $q_y < n_z$ can be related to the existence of waveguide modes in a dielectric slab. Indeed, as for standard waveguide modes, their wavevector diminishes when reducing the film thickness[25]. Instead, along the [100] direction the transfer matrix calculations predict two types of PPs modes. There are modes existing only for $q_x > n_z$, which have increasing wavevector with decreasing thickness, analogously to what has been reported for bulk (i.e., volume-confined) mid-IR phonon polaritons in hBN[26] and $MoO_3$[27] thin films. Then, for thicknesses below $\approx 200$ nm, we observe the emergence of a lower-order mode with a different character, since it exists within the $MoOCl_2$ light-cone, i.e., for $q_x < n_z$, approaching the substrate light-cone at $q_x = n_{sub}$ for large thicknesses.

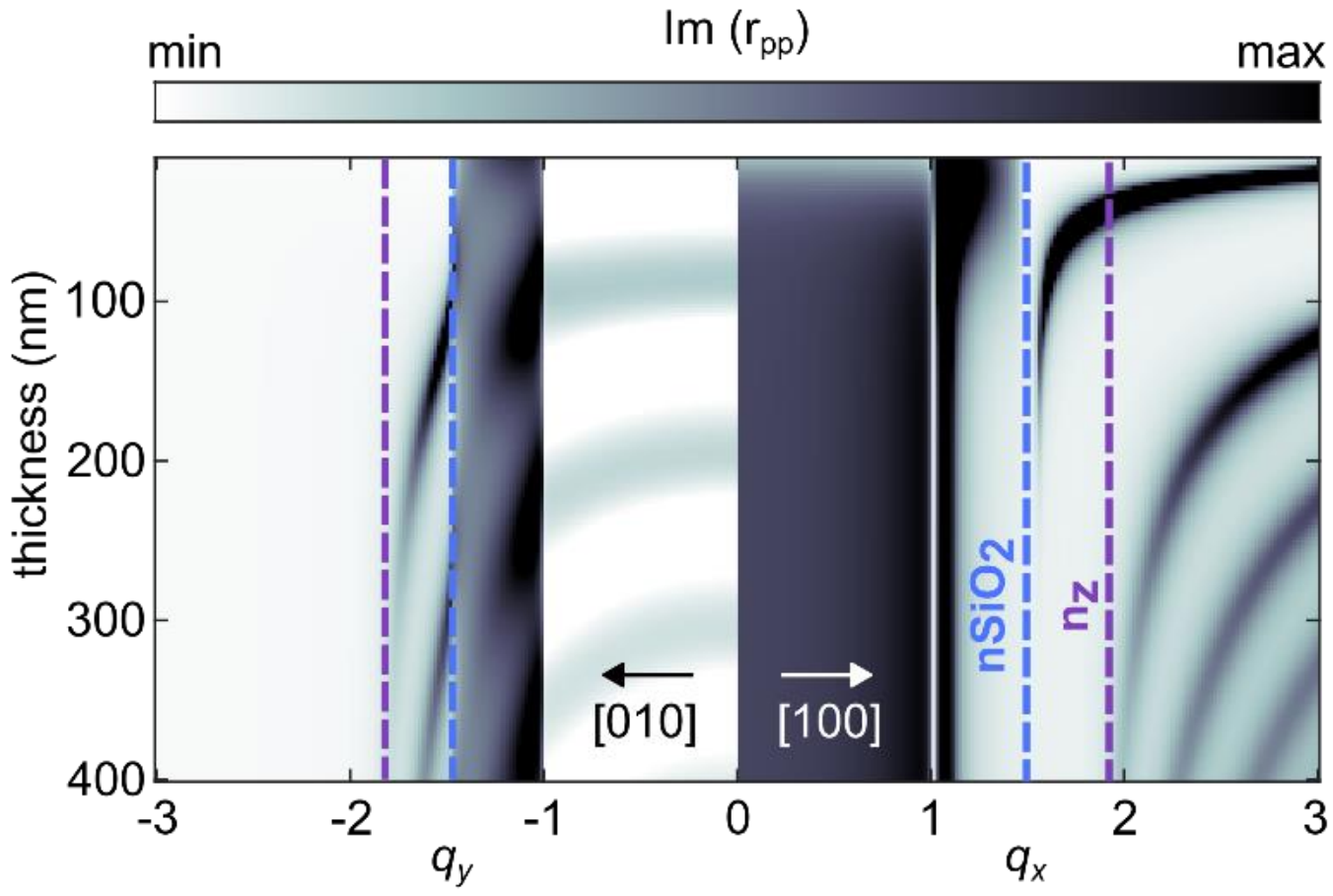

**Figure S15. Transfer matrix calculated propagating modes.** Transfer matrix calculation of the imaginary part of the reflection coefficient $\mathrm{Im}(r_{pp})$ as a function of the flake thickness in the [010] and [100] crystal directions at $\lambda_0 = 633$ nm. Purple and blue dashed lines indicate the values of $n_z = 1.9$ and $n_{sub} = 1.5$, respectively.

### Electric field distribution in thin films

Additional insights on the modes properties as calculated from the transfer matrix method can be obtained by plotting the field distribution as shown in Fig. S16. The out of plane component of the field $E_z$ as a function of the in-plane wavevector at $\lambda_0 = 633$ nm and for a film thickness of 80 nm is investigated here. The circles indicate respectively the refractive index of air (white, $n_{air} = 1$), of the $SiO_2$ substrate (blue, $n_{sub} = 1.5$) and the out of plane refractive index of $MoOCl_2$ (purple, $n_z = 1.9$). Within the blue circle, a high signal associated with propagating waves is observed. The field related with the lenticular ($|q| < n_z$) and for the first two hyperbolic modes can be seen in Fig. S16a. Projections of the field distribution along the stack vertical direction are shown in Fig. S16b, c for the $q_y = 0$ direction and by the same line tilted clockwise by 7° as indicated by the dashed lines in panel a. At high $|q|$ a bulk waveguide-like mode with a characteristic maximum inside the film can be observed in both Fig. S16b, c. Vertical dashed lines indicate the $n_{air}$, $n_{sub}$ and $n_z$ corresponding to the concentric rings in panel a. This corresponds to the modes observed in Fig. S15, which only exist for $|q| > n_z$. By extending the wavevector range of the plot, higher order modes with increasing number of field nodes in the $MoOCl_2$ film can be revealed. In the 7° cut in Fig. S15c another mode appears, at intermediate $|q|$ between the lenticular and the waveguide mode, corresponding to the first hyperbolic mode that is only intersected with a tilted cut as shown in panel a. The field distribution of the lenticular and first hyperbolic modes show similar features, with a single field node in the film and high field concentration at the $MoOCl_2$ boundaries. Therefore, considering such $E_z$ distributions, we can conclude that the PPs associated to the lenticular and lowest order hyperbolic modes are surface polariton modes, distinct to the bulk hyperbolic modes existing only for $q_x > n_z$, characterized by the highest field region located inside the flake volume.

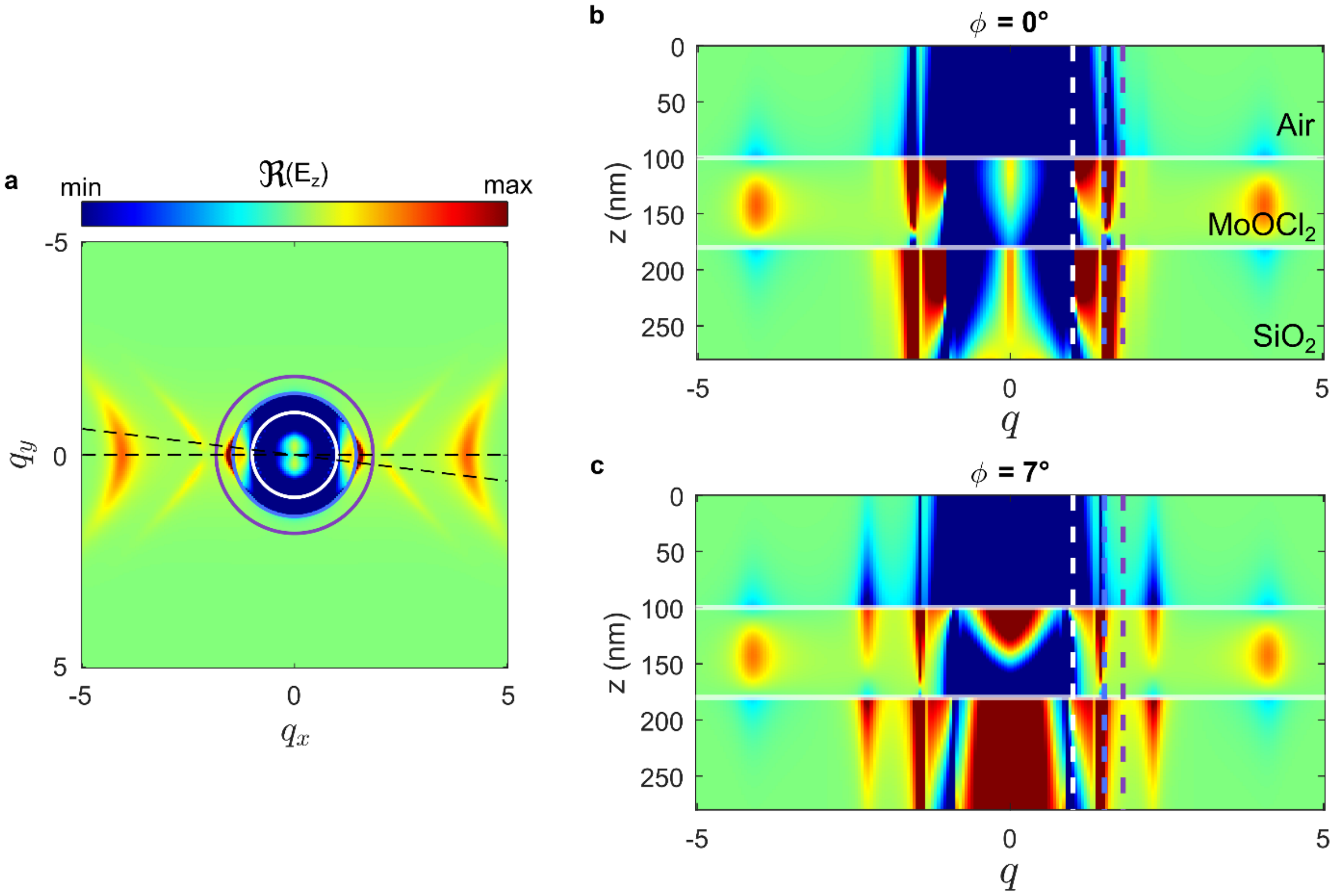

**Figure S16. Transfer matrix calculated field. a** $E_z$ field calculated from transfer matrix at $\lambda_0$ = 633 nm for an 80 nm flake on $SiO_2$. The field is taken just below the $MoOCl_2$ top surface. Vertical cuts of $E_z$ field along the horizontal **b** and tilted by 7° **c** dashed lines shown in panel a.

These plots highlight the connection between the lenticular and first hyperbolic modes, which appear to be surface modes as the maximum field is achieved at the material boundaries.
In Fig. S17 we show a similar plot, but we report instead the 1-dimensional projection of the absolute value of the field $|E_z|$ for three selected points in the $(q_x, q_y)$ plane as highlighted by the arrows in Fig. S17a. The $|E_z|$ profiles in Fig. S17b-d are representative of the lenticular, first and second order hyperbolic modes. Again, the similarity between the lenticular and first hyperbolic modes is confirmed, as they display a similar field distribution.

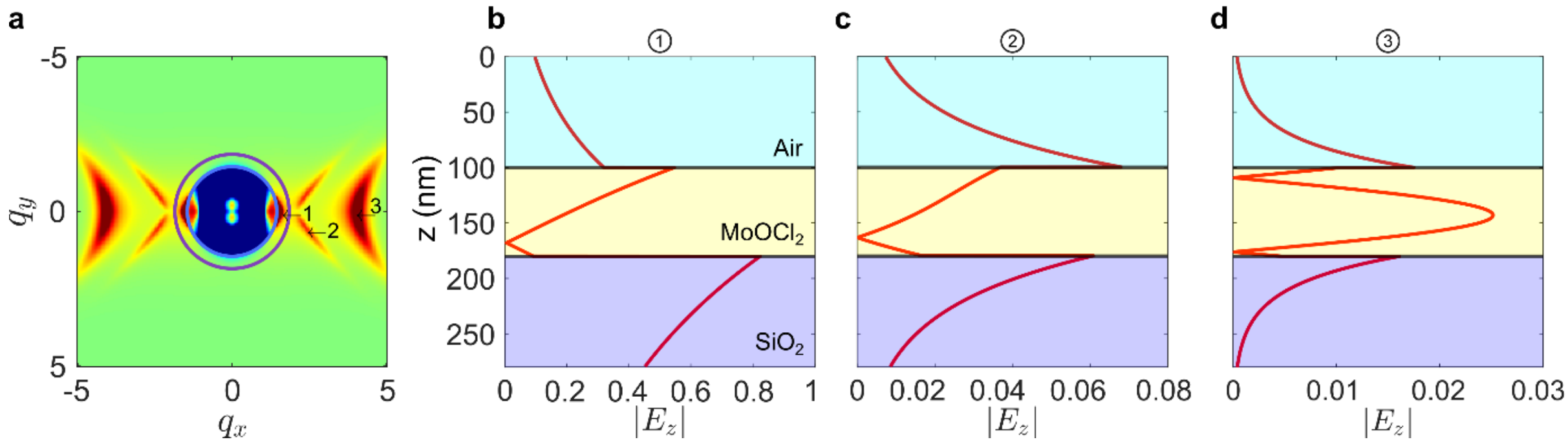

**Figure S17. 1D-projections of the calculated field. a** $E_z$ field calculated from the transfer matrix at $\lambda_0$ = 633 nm for a 80 nm flake on $SiO_2$. The field is taken just below the $MoOCl_2$ top surface. The arrows indicate the points where the 1D projections shown in **b-d** are taken.

## S6 – Details of the disk launcher fabrication

We use a gold disk fabricated through standard electron beam lithography to launch hyperbolic PPs in a $MoOCl_2$ flake as shown in Fig. 4 of the main text. A scanning electron microscope (SEM) image of the 42 nm-thick flake with the disk on top is shown in Fig. S18a. Part of the resist used for the lithography got exposed during electron irradiation, forming the square base shown in the s-SNOM image in Fig. S18, which is covered by the yellow dot in the main text. The base obscures the visibility of the first PPs fringe but does not significantly alter their excitation from the gold disk. In Fig. S18c we show an edge of the flake with the disk fabricated on top (the bar is a marker used for the fabrication alignment), confirming the 42 nm thickness as evaluated from the profile in Fig. S18d.

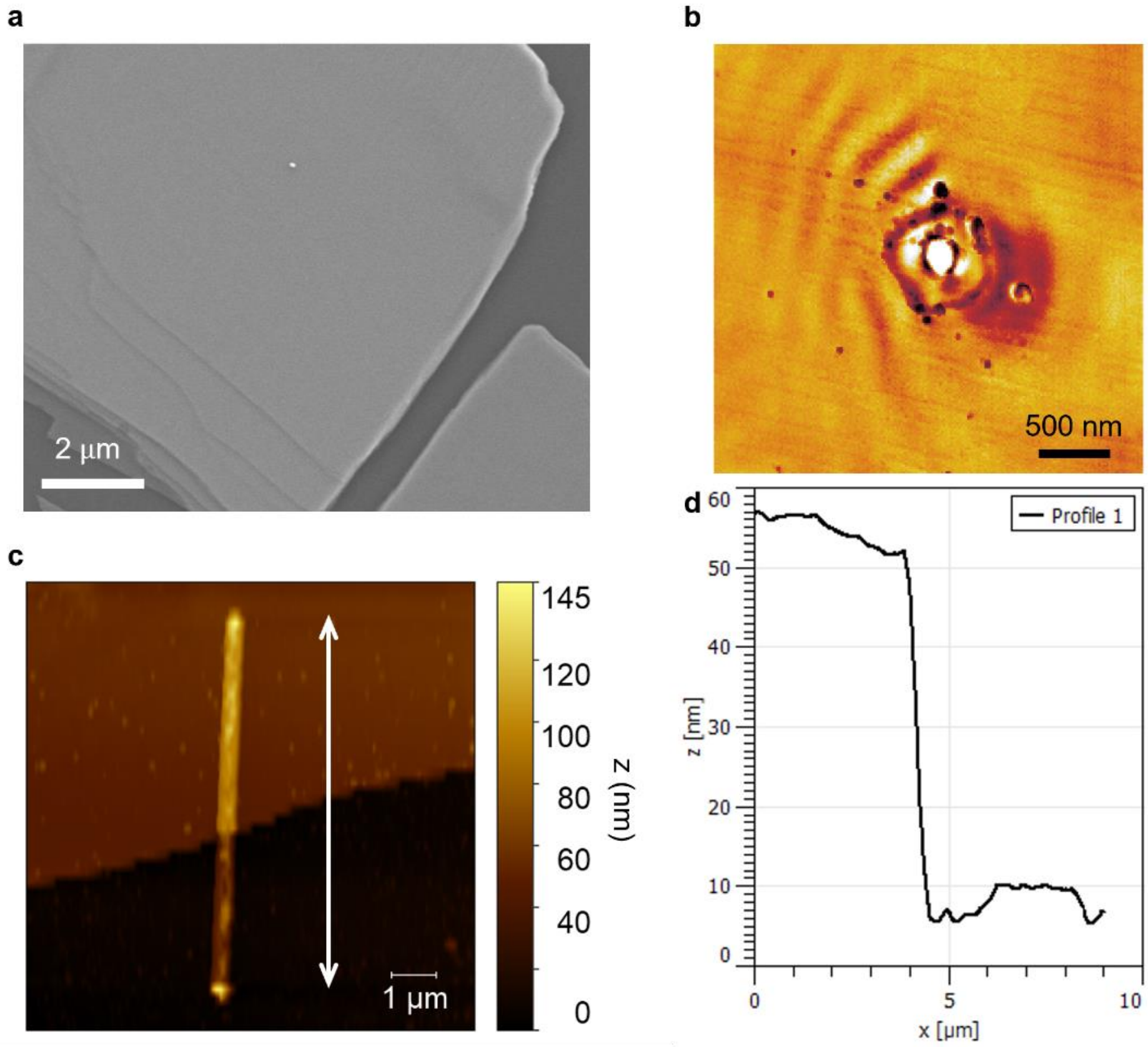

**Figure S18. Disk launcher on the 42 nm flake. a** SEM image of a portion of the flake with the gold disk fabricated on top. **b** s-SNOM image at $\lambda_0 = 532$ nm without covering the launcher. The square base is a result of fabrication imperfection. **c** AFM map an edge of the 42 nm flake. The gold bar is a marker used for fabrication. **d** Topography profile extracted along the white line in panel **c**.

## S7 - Computation of the FFT maps from the experimental data

To calculate the FFT maps shown in Fig. 4 of the main text from the experimental data we use the procedure outlined below. As the IFCs are symmetric in the $x$ and $y$ axis, we only select a region of the s-SNOM image (Fig. S19a) and use it to produce a real-space image with the correct symmetry by flipping it along the horizontal and vertical directions (Fig. S19b). In this way we can get rid of the asymmetries due to imperfections in the launcher fabrications and the difference in periodicities on the right and left of the disk due to the changing relative direction of the free-space and polariton wavevectors. After this step we remove a constant background from the complex map, we apply a windowing function and pad the image with zeros. The absolute value of the FFT of the image treated in this way is shown in Fig. S19c. As final step, we shift the $x$ values of the FFT by $\pm\kappa_0 \cos\,\phi$ where $\phi = 30°$ is the incident angle of the light focused by the parabolic mirror in the s-SNOM experiments (Fig. S19d). This step corrects the disk-launched fringes periodicity and reveal the true polariton wavelength along the $x$-axis[28]. While this procedure corrects the fringes along $x$, it does not consider that the measured periodicity depends on the relative angle between the wavevector and the normal direction to the finite-size disk edge[29], which changes along the perimeter of the disk launcher and should introduce additional (smaller) corrections in the maps in Fig. 4.

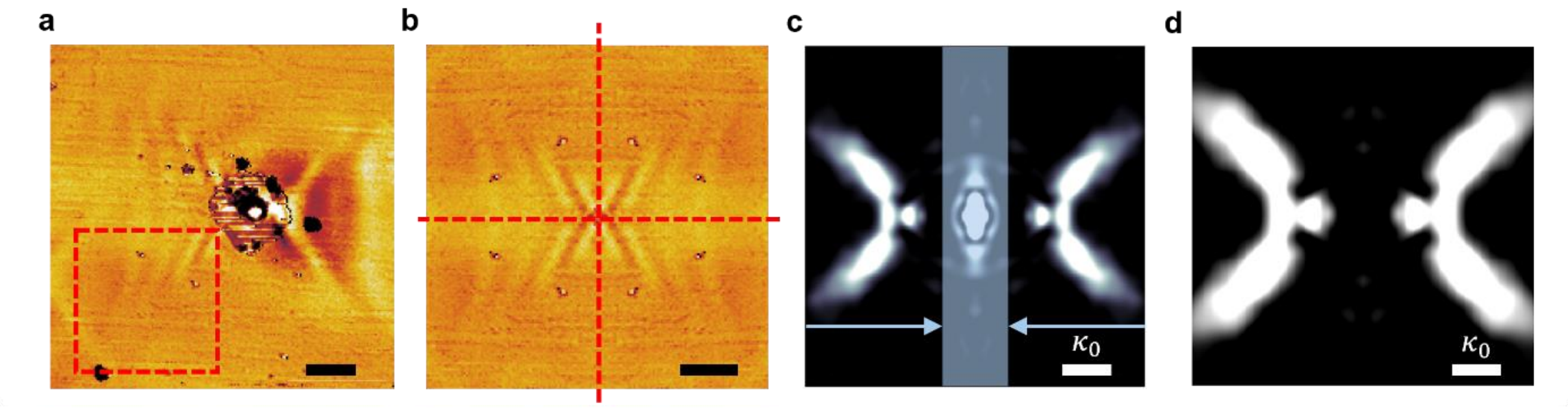

**Figure S19. Calculation of the experimental FFT maps. a** s-SNOM image at $\lambda_0$ = 730 nm on the 42 nm flake shown in Fig. 4 of the main text. The dashed red box indicates the region used to compute the FFT map. **b** Symmetrized image obtained from the box in **a** after flipping in along the vertical and horizontal directions. **c** |FFT| map of **b** after removing a constant background, multiplying it by a windowing function and zero padding it. **d** |FFT| map after correcting for the $\pm\kappa_0 \cos\,\phi$ shift in the experimental data due to the relative direction of the polariton and free space momenta.

## S8 - Effect of the near-field source size on the simulated FFT maps

The FFT maps computed from CST simulations (Fig. S20a-d) confirm that the hyperbolic modes are mainly excited for longer wavelengths, while the lenticular PPs assume considerable intensity only below 633 nm. While the spectra of probed wavevectors in the simulations depends on the geometry of the near-field source (the distance from the surface and its size, as discussed below), the comparison of the relative behaviour at different wavelengths with the same source can still give useful information to interpret the experimental results. As seen from Fig. S20, numerical calculations agree quite well with the experimental results of Fig. 4 of the main text.

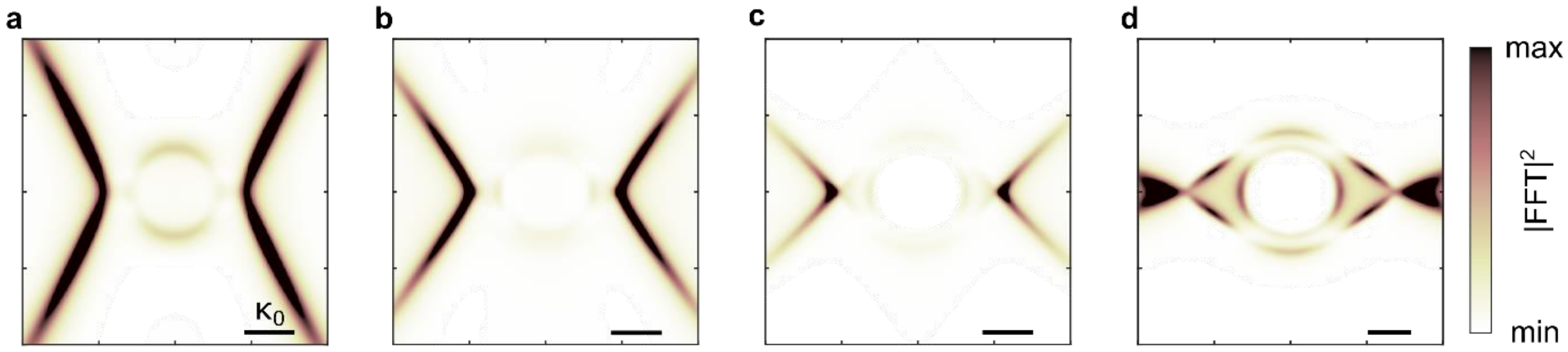

**Figure S20. FFT maps of numerical CST simulations as a function of $\lambda_0$. a** 830 nm. **b** 730 nm. **c** 633 nm. **d** 532 nm. Source size: length $L = 50$ nm, radius $r = 10$ nm.

Regarding the near-field source for CST simulations, as explained in Methods, we use a perfect electric conductor (PEC) cylinder with rounded caps (Fig. S21a) excited by a discrete port. The physical dimensions of the source determine the wavevectors exciting the polaritons in the $MoOCl_2$ below. Generally, the larger the source, the smaller the wavevectors associated with it. Therefore, the FFT map shown in the main text depends on the dimension of the source. In Fig. S21b-d we report the FFT maps obtained from a set of simulations ($\lambda_0$ = 633 nm) where we changed the curvature radius and the length of the source. As expected, increasing the source dimensions decreases the intensity in the FFT maps of the hyperbolic high wavevectors components while increasing the weight of the lenticular and circular IFCs. A similar effect can be expected when changing the disk radius and thickness used in the experiments to launch the plasmon polaritons.

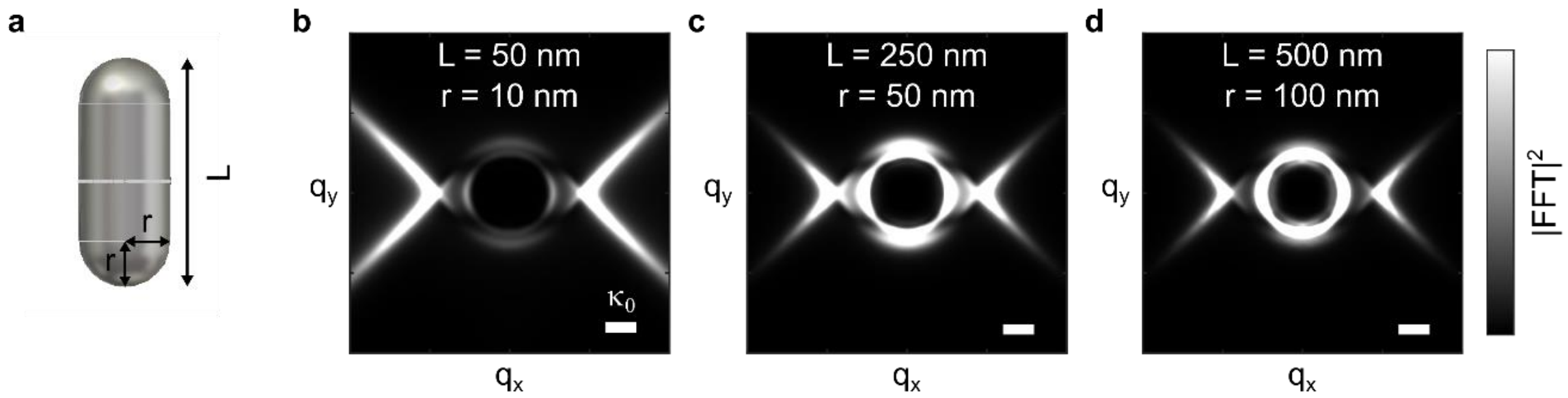

**Figure S21. Dependence of the FFT maps on the near field source dimension. a** Sketch of the source used in the CST simulations. A PEC cylinder of length $L$ and radius $r$ is excited by a discrete port. **b-d** FFT maps obtained from simulations ($\lambda_0$ = 633 nm) when changing $L$ and $r$ of the near field source.

## S9 – Analysis of the propagation

### Extraction of the propagation angle

We extracted the maximum propagation angle on Gwyddion, by taking linecuts starting from the centre of the launcher along the smallest angle at which we stop observing oscillations, as shown in Fig. S22a. The total aperture angle $2\alpha_{\mathrm{H}}$ is then extracted as the difference between the angles such linecuts form with the positive horizontal axis. This procedure was applied to both the left (red lines) and right (blue lines) fringes, and the value reported in Fig. 4m of the main text are the average of the two numbers.
Since the two computations give very similar results but the procedure depends on the operator, we give an estimation of the error bar as half the width of the observed fringes, propagated on the difference.

### Fitting of polariton decay length.

We extract the PPs profiles launched by the gold disk shown in Fig. 4 of the main text with the Gwyddion linecuts at an angle slightly lower than $\alpha_{\mathrm{H}}$, where we can clearly see the oscillations in the signal (see dashed line in Fig S22a and Fig. 4d in the main text).
To fit the profiles, we use a decaying exponential function defined as:

$$S_3 = Ae^{-Im(q)x}\sin(Re(q)x + \phi) + B,$$

where $A, B, \phi$ and $q$ are the fitting parameters. This allows to find the quality factor as $Q_p = \mathrm{Re}(q)/\mathrm{Im}(q)$ $(= 2\pi L/\lambda_0$, where $L$ is the propagation length). However, this procedure underestimates the decay length as it neglects the intensity decay due to geometrical spreading of the wavefronts, independently from material absorption. For spherical waves this effect can be accounted for by dividing with $\sqrt{x}$ so that the fitting function becomes:

$$S'_3 = Ae^{-Im(q)x}\sin(Re(q)x + \phi)/\sqrt{x} + B$$

Fitting with $S'_3$ yields higher propagating $Q_p$ as shown in Fig. S22b. For simplicity, instead of fitting with $S'_3$, we correct the experimental data by dividing them by $\sqrt{x}$ to then fit them with expression $S_3$. While the intensity of spherical waves decays as $\sqrt{x}$ in the absence of losses, it is not straightforward to determine which is the geometrical decay factor for modes with hyperbolic and lenticular shape. Therefore, we prefer to report in the main text the lower boundary for $Q_p$ obtained by fitting with expression $S_3$.
Interestingly, the PPs propagation length has an anomalous increase at 633 nm (Fig. 4l and Fig. S22b), which we associate with the transition from the hyperbolic to the lenticular mode.

### Lifetime of the polaritons

With the knowledge of $Q_p$, it is possible to estimate a value for the hyperbolic PPs lifetime $\tau_p$, defined as:

$$\tau_p := \frac{L_p}{v_g} = \left(\frac{Q_p}{2\pi}\lambda_0\right)\frac{1}{v_g},$$

where $v_g$ is the group velocity, which can be computed from the theoretical PPs dispersion ($\omega$-$k$) as:

$$v_g := \frac{d\omega}{dk} = \frac{1}{\hbar}\frac{dE}{dk}.$$

In general, $v_g$ is a function of the in-plane direction $\beta$ in momentum space (see Fig. S8 for the definition) and $\lambda_0$. For the $\tau_p$ extraction, we are interested in the group velocity along the direction in the crystal defined by the propagation of the hyperbolic PPs. In pratice, we start by computing the PPs dispersion (in the thin film approximation) as a function of $\beta$, as shown in Fig. S22c. We then compute the $d/dk$ derivative at the same wavelengths at which $Q_p$ is measured (dahsed lines in Fig. S22c; 532 nm is excluded since there we are mainly observing lenticular PPs). Finally, we look at the $v_g$ at the specific $\beta_0$ corresponding to the direction (in real space, see dashed line in Fig. S22a) of the line profiles shown in Fig. 4l (or, equivalently, Fig. S22b) from which $Q_p$ is obtained. Specifically, $\beta_0$ is complementary to that direction, since we are in the asimptotic limit and the Poynting vector is orthogonal to the wavevector. Notably, $\beta_0$ changes with $\lambda_0$ along with the aperture of the hyperbolic rays (see Fig. 4n). The values of $\beta_0$ at $\lambda_0 = 633, 730, 830$ nm are marked by black dots in Fig. S22d, where we plot $\tau_p$ as a function of $\beta$. For those points, the derived $v_g(\beta_0)$ is on the order of $10^5 - 10^6$ m/s and the corresponding $\tau_p(\beta_0)$ ranges from $120$ fs (633 nm) to $190$ fs (830 nm). These values are an order of magnitude smaller compared to the ones observed for low-loss PhPs at mid-IR frequencies. However, if we consider that the frequency of light is one order of magnitude larger in the visible range, the normalized quantity $\tau_p/T_0$ - where $T_0$ is the period of free-space light (~1-2 fs in the visible vs. ~10-100 fs in the mid-IR) - is in the same order of magnitude of PhPs. This explains why even if the lifetime is smaller, the Q factor of the propagating mode is similar for mid-IR PhPs and visible PPs in $MoOCl_2$.

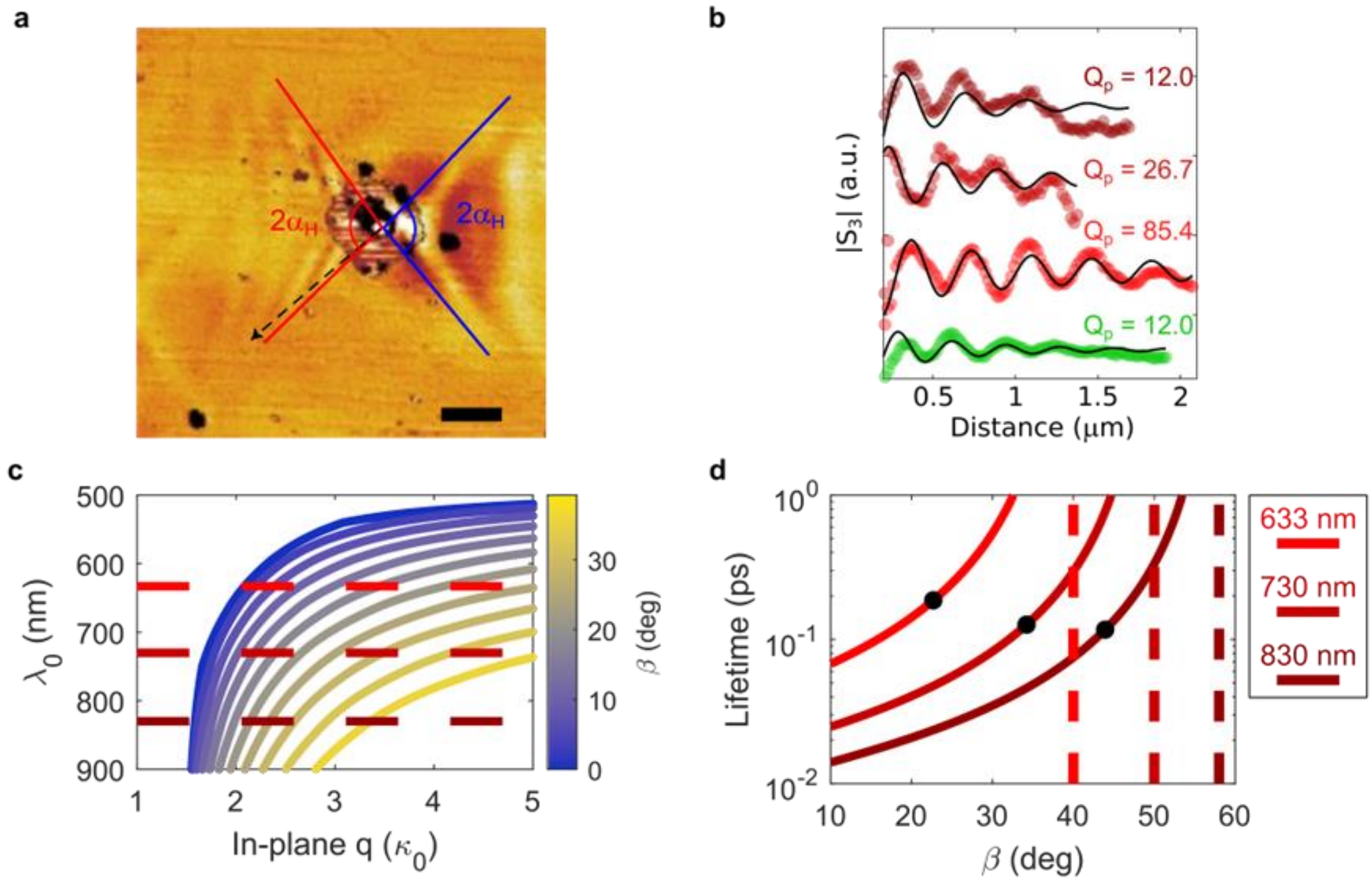

**Figure S22. Propagation properties of PPs. a** Extraction of the aperture angle for the propagation in real space. Black arrow indicates the direction of the extracted PPs profiles ($\sim 10°$ below $\alpha_H$). **b** Fit of the PPs profiles launched by the gold disk shown in Fig. 4l of the main text. We fit here the data divided by $\sqrt{x}$ to consider the intensity decay due to geometrical spreading of the wavefronts. Consequently, the extracted $Q_p$ are larger than the ones reported in Fig. 4l. **c** Dispersion relation of PPs for a thin film computed for different in-plane angles $\beta$ (in momentum space). The horizontal dashed line indicates the wavelength at which we extracted the lifetime (633 nm, 730 nm and 830 nm). **d** Lifetime of the PPs as a function of $\beta$. The curves are computed using the $Q_p$ reported in Fig. 4l of the main text and $v_g$ extracted for the three wavelengths highlighted in **c.** Black dots are in correspondence of $\beta_0$. Vertical dashed lines mark the asymptotic propagation angles ($\beta = \pi/2 - \alpha_H$).

For a fair comparison of the lifetime with well known plasmonic materials, we should consider it at a comparable degree of confinement (i.e., the normalized wavevector $q$), as for surface PPs a larger $q$ inevitably leads to higher losses due to the increased field penetration in the material.[30] For this, we consider PPs in a $40$ nm gold film. At a wavelength of $532$ nm, we obtain a lifetime of 32 fs (as extracted from the propagation lengths taken from Ref. [31]) at $q = 2$, while for MoOCl2 at $830$ nm we retrieve a $190$ fs lifetime for an even greater confinement $q = 3.3$, showcasing the small lossess in our platform.

**S10 - Dispersion along the metallic direction in a 20 nm flake.**

To verify the predicted PPs dispersion in the metallic direction, we perform s-SNOM nano-imaging of polaritons launched by the edge of a 20 nm flake, with the topography shown in Fig. S23a. The flake has a straight edge (on the right) corresponding to the [010] crystal direction. The corresponding s-SNOM amplitude map (scanning along the [100] direction) at 800 nm shows periodic oscillations from both flake edges (Fig. S23b). We chose the right edge for our investigation as it is straight, ensuring the correspondence to one of the crystal lattices. Exemplary s-SNOM maps of a smaller $4 \times 5$ μm$^2$ region at 720, 800 and 880 nm are shown in Fig. S23c-d respectively. The periodic oscillations are analysed by averaging each profile along the vertical direction, followed by extraction of the position of the maximum of their Fourier transform (Fig. S23f). The associated error is estimated as half of the full width at half maximum of the peak. The experimental values are plotted on top of the calculated transfer matrix as shown in Fig. S23g. The good agreement obtained by dividing by 2 the experimental momentum implies we are measuring the tip-launched contribution (i.e., PPs launched by the tip, reflected from the edge and scattered back to the detector by the tip), which appears as oscillations of periodicity that is half of the polariton wavelength[29] (so that the measured momentum is doubled). The reason why we are observing only tip-related (and not edge-related) contributions can be motivated by the geometry of the illumination: indeed, laser light is coming from the left side in the displayed images, hence the tip "shadows" the direct illumination of the right edge, hindering the excitation of the edge-component. This is in line with the observed fringes on the left edge, instead, which are characterized by a different periodicity, which changes with the orientation of the flake with respect to the illumination direction (not shown here).

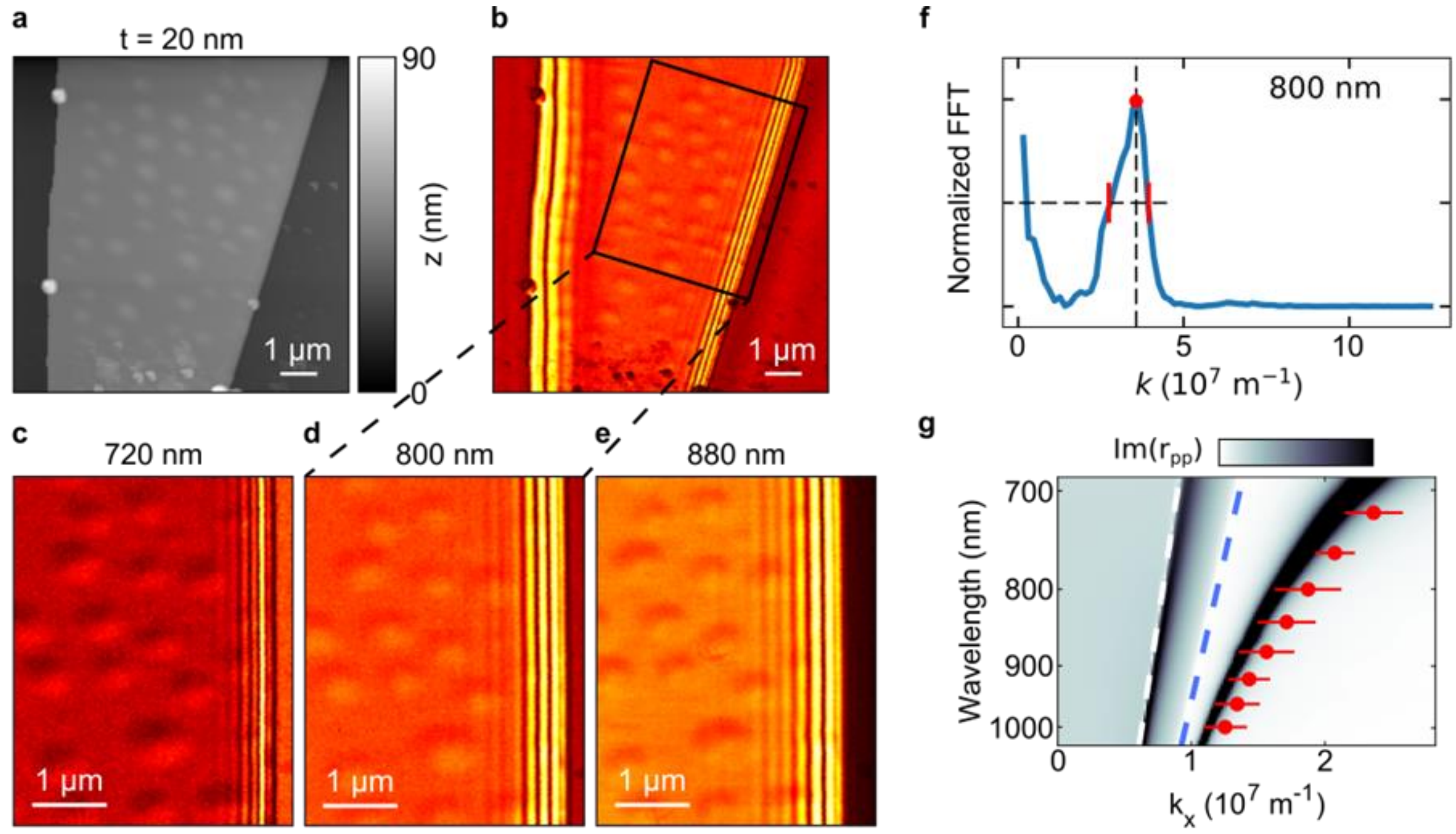

**Figure S23. Polariton dispersion for a 20 nm flake. a** Topography and fourth order demodulate s-SNOM amplitude **b** of a 20 nm $MoOCl_2$ flake. **c-d** Exemplary s-SNOM maps at $\lambda_0$ = 720, 800 and 880 nm. **f** The polariton momentum is extracted as the position of the maximum of the FFT of the vertically averaged profile (red dot). The errors are estimated as half of the full width at half maximum of the peak (red lines). The position of the peaks in momentum space extracted from the FFT are plotted as red dots in panel **g** on top of the corresponding transfer matrix calculation along the $x$ direction ([100] crystal direction). The experimental momentum (and the corresponding error) is divided by 2, indicating that we are measuring the tip-launched component. Dashed lines indicate the air (white) and $SiO_2$ (blue) light lines.

## References

1. Gao, H., Ding, C., Sun, L., Ma, X. & Zhao, M. Robust broadband directional plasmons in a MoOCl2 monolayer. *Phys Rev B* **104**, 205424 (2021).

2. Zhao, J. *et al.* Highly anisotropic two-dimensional metal in monolayer MoOCl2. *Phys Rev B* **102**, (2020).

3. Korzeb, K. *et al.* Compendium of natural hyperbolic materials. *Optics Express, Vol. 23, Issue 20, pp. 25406-25424* **23**, 25406–25424 (2015).

4. Esslinger, M. *et al.* Tetradymites as Natural Hyperbolic Materials for the Near-Infrared to Visible. *ACS Photonics* **1**, 1285–1289 (2014).

5. Zhao, M., Li, W., Gao, H. & Zhang, X. Tunable broadband hyperbolic light dispersion in metal diborides. *Optics Express, Vol. 27, Issue 25, pp. 36911-36922* **27**, 36911–36922 (2019).

6. Ruta, F. L. *et al.* Hyperbolic exciton polaritons in a van der Waals magnet. *Nature Communications 2023 14:1* **14**, 1–9 (2023).

7. Lee, Y. U. *et al.* Organic Monolithic Natural Hyperbolic Material. *ACS Photonics* **6**, 1681–1689 (2019).

8. Lee, Y. U., Yim, K., Bopp, S. E., Zhao, J. & Liu, Z. Low-Loss Organic Hyperbolic Materials in the Visible Spectral Range: A Joint Experimental and First-Principles Study. *Advanced Materials* **32**, 2002387 (2020).

9. Caldwell, J. D. *et al.* Sub-diffractional volume-confined polaritons in the natural hyperbolic material hexagonal boron nitride. *Nat Commun* **5**, (2014).

10. Ma, W. *et al.* Ghost hyperbolic surface polaritons in bulk anisotropic crystals. *Nature* **596**, 362–366 (2021).

11. Hu, G. *et al.* Real-space nanoimaging of hyperbolic shear polaritons in a monoclinic crystal. *Nat Nanotechnol* **18**, 64–70 (2023).

12. Zheng, Z. *et al.* A mid-infrared biaxial hyperbolic van der Waals crystal. *Sci Adv* **5**, (2019).

13. Álvarez-Pérez, G., Voronin, K. V., Volkov, V. S., Alonso-González, P. & Nikitin, A. Y. Analytical approximations for the dispersion of electromagnetic modes in slabs of biaxial crystals. *Phys Rev B* **100**, 235408 (2019).

14. Agranovich, V. M. (Vladimir M. & Ginzburg, V. L. (Vitaliĭ L. *Crystal Optics with Spatial Dispersion, and Excitons*. (Springer-Verlag, 1984).

15. Narimanov, E. E. Dyakonov waves in biaxial anisotropic crystals. *Phys Rev A (Coll Park)* **98**, 013818 (2018).

16. Álvarez-Pérez, G., Voronin, K. V., Volkov, V. S., Alonso-González, P. & Nikitin, A. Y. Analytical approximations for the dispersion of electromagnetic modes in slabs of biaxial crystals. *Phys Rev B* **100**, 235408 (2019).

17. D'yakonov, M. ~I. New type of electromagnetic wave propagating at an interface. *Soviet Journal of Experimental and Theoretical Physics* **67**, 714 (1988).

18. Ni, X. *et al.* Observation of directional leaky polaritons at anisotropic crystal interfaces. *Nat Commun* **14**, (2023).

19. Narimanov, E. E. Dyakonov waves in biaxial anisotropic crystals. *Phys Rev A (Coll Park)* **98**, 013818 (2018).

20. Feng, Y. *et al.* Visible to mid-infrared giant in-plane optical anisotropy in ternary van der Waals crystals. *Nature Communications 2023 14:1* **14**, 1–8 (2023).

21. Chiodini, S. *et al.* Angstrom-Resolved Metal-Organic Framework-Liquid Interfaces. *Scientific Reports 2017 7:1* **7**, 1–6 (2017).

22. Gan, Y. Atomic and subnanometer resolution in ambient conditions by atomic force microscopy. *Surf Sci Rep* **64**, 99–121 (2009).

23. Wang, Z. *et al.* Fermi liquid behavior and colossal magnetoresistance in layered MoOC l2. *Phys Rev Mater* **4**, (2020).

24. Paarmann, A. & Passler, N. C. Generalized 4 × 4 matrix formalism for light propagation in anisotropic stratified media: study of surface phonon polaritons in polar dielectric heterostructures. *JOSA B, Vol. 34, Issue 10, pp. 2128-2139* **34**, 2128–2139 (2017).

25. Hu, F. & Fei, Z. Recent Progress on Exciton Polaritons in Layered Transition-Metal Dichalcogenides. *Adv Opt Mater* **8**, 1901003 (2020).

26. Dai, S. *et al.* Tunable phonon polaritons in atomically thin van der Waals crystals of boron nitride. *Science (1979)* **343**, 1125–1129 (2014).

27. Ma, W. *et al.* In-plane anisotropic and ultra-low-loss polaritons in a natural van der Waals crystal. *Nature 2018 562:7728* **562**, 557–562 (2018).

28. Ni, X. *et al.* Observation of directional leaky polaritons at anisotropic crystal interfaces. *Nat Commun* **14**, (2023).

29. Mancini, A. *et al.* Near-Field Retrieval of the Surface Phonon Polariton Dispersion in Free-Standing Silicon Carbide Thin Films. *ACS Photonics* **9**, 3696–3704 (2022).

30. Maier, S. A. *Plasmonics : Fundamentals and Applications*. (Springer, 2007).

31. Lebsir, Y., Boroviks, S., Thomaschewski, M., Bozhevolnyi, S. I. & Zenin, V. A. Ultimate Limit for Optical Losses in Gold, Revealed by Quantitative Near-Field Microscopy. *Nano Lett* **22**, 5759–5764 (2022).